%% file: journal.tex
\documentclass[preprint,12pt]{elsarticle}

\usepackage[english]{babel}
\usepackage{graphicx,url}       % pacote para importar figuras
\usepackage{mathptmx}           % p/ usar fonte Adobe Times
\usepackage{subcaption}
\usepackage[dvipsnames,svgnames,table]{xcolor}
\usepackage{hhline}
\usepackage{etoolbox}
\usepackage{cellspace}
\usepackage{rotating}

\usepackage{verbatim}

\usepackage{float}
\usepackage[notransparent]{svg}
\usepackage{array}
\newcolumntype{P}[1]{>{\centering\arraybackslash}p{#1}}
\usepackage{pdfpages}
\usepackage{url}
\usepackage{color}
\usepackage{xspace}

\usepackage{listings}
\usepackage{tabularx}
\newcolumntype{Y}{>{\centering\arraybackslash}X}
\usepackage{pbox}

\usepackage{multirow} 
\usepackage{setspace}
\usepackage{amsmath}
\usepackage{amssymb}
\usepackage{amsfonts}
\usepackage{amsthm}

\usepackage[utf8]{inputenc}
\usepackage[T1]{fontenc}
\usepackage{algpseudocode,algorithm,algorithmicx}
\usepackage{newtxmath}
\definecolor{MyBlue}{HTML}{5f8dd3}

\newcommand{\blue}[1]{\textcolor{blue}{#1}}

\definecolor{lightgray}{gray}{0.9}
\definecolor{darkgray}{gray}{0.5}

\usepackage[nolist,nohyperlinks]{acronym}
\usepackage{mfirstuc}
\usepackage{hyperref}
\hypersetup{
	draft=false,
	colorlinks=false,
	linktocpage=true,
	pdfstartview=FitH,
	breaklinks=true,
	pdfpagemode=UseNone,
	pageanchor=true,
	pdfpagemode=UseOutlines,
	plainpages=false,
	bookmarksnumbered,
	bookmarksopen=false,
	bookmarksopenlevel=1,
	hypertexnames=true,
	pdfhighlight=/O,
	urlcolor=Maroon,linkcolor=MyBlue,citecolor=DarkGreen, % <--- for screen
	pdftitle={},
	pdfauthor={},
	pdfsubject={},
	pdfkeywords={},
	pdfcreator={pdfLaTeX},
	pdfproducer={LaTeX with hyperref}
}

\usepackage[textwidth=30mm]{todonotes}
\usepackage{soul}
\setstcolor{red}
\sethlcolor{SkyBlue}

\theoremstyle{plain}

\newtheorem*{corollary*}{Corollary}

\theoremstyle{definition}

\newtheorem*{definition*}{Definition}

\theoremstyle{remark}

\newtheorem*{remark*}{Remark}

\newcommand{\flow}{s}

\newcommand{\flowset}{\mathcal{S}}

\newcommand{\flowid}{k}

\newcommand{\flowidset}{\mathcal{K}}

\newcommand{\feature}{a}

\newcommand{\alphabet}{\Omega}

\newcommand{\field}{h}

\newcommand{\featurei}{i}
\newcommand{\featurej}{j}

\newcommand{\featureids}{\mathrm{N}}

\newcommand{\coup}{e}

\newcommand{\ie}{\textit{i.e., }}                   %i.e.,
\newcommand{\eg}{\textit{e.g., }}                   %e.g., 
\newcommand{\etal}{\textit{et al. }}               %Et al.
\newcommand{\node}{\eta}

\newcommand{\1}{(\textit{i})}
\newcommand{\2}{(\textit{ii})}

\begin{acronym}
	\acro{ANN}{Artificial Neural Network}
	\acro{MLP}{Multi-Layer Perceptron}
	\acro{NIDS}{Network Intrusion Detection System}
	\acro{IDS}{Intrusion Detection System}
	\acro{S-NIDS}{Signature-based Network Intrusion Detection System}
	\acro{A-NIDS}{Anomaly-based Network Intrusion Detection System}
	\acro{NID}{Network Intrusion Detection}
	\acro{EFC}{Energy-based Flow Classifier}
	\acro{DoS}{Denial of Service}
	\acro{ML}{Machine Learning}
	\acro{Ada B.}{Adaboost}
	\acro{DT}{Decision Tree}
	\acro{LR}{Linear Regression}
	\acro{NB}{Naive Bayes}
	\acro{KNN}{K-Nearest Neighbors}
	\acro{RF}{Random Forest}
	\acro{SVM}{Support Vector Machine}
	\acro{SVC}{Support Vector Classifier}
	\acro{IoT}{Internet of Things}
	\acro{SDN}{Software Defined Networks}
	\acro{CNN}{Convolutional Neural Network}
	\acro{PCA}{Principal Component Analysis}
	\acro{DBN}{Deep Belief Network}
	\acro{LSTM}{Long Short-term Memory}
	\acro{ELM}{Extreme Learning Machine}
	\acro{HMLD}{Hybrid Multi-Level Data Mining}
	\acro{AUROC}{Area Under the Receiver Operating Characteristics}
	\acro{IoMT}{Internet of Medical Things}
\end{acronym}

\journal{\begin{minipage}{0.5\textwidth}Computer \& Security\hfill \\\tiny CFP: Advances in Robust Intrusion Detection through Machine Learning\end{minipage}}

\begin{document}
	
	\title{A Novel Open Set Energy-based Flow Classifier for Network Intrusion Detection}
	
	\author[UnB]{Manuela~M.~C.~Souza}
	\ead{manuela.matos@aluno.unb.br}
	\author[BSC]{Camila~T.~Pontes}
	\ead{camila.pontes@bsc.es}
	\author[UnB]{Jo\~ao~J.~C.~Gondim}
	\ead{gondim@unb.br}
	\author[UnB]{Lu\'is~P.~F.~Garcia}
	\ead{luis.garcia@unb.br}
	\author[VT]{Luiz DaSilva}
	\ead{ldasilva@vt.edu}
	\author[UnB]{Eduardo F. M. Cavalcante}
	\ead{202006368@aluno.unb.br}
	\author[UnB]{ Marcelo~A.~Marotta$^*$}
	\ead{marcelo.marotta@unb.br}

	\affiliation[UnB]{organization={University of Brasilia},%Department and Organization
		addressline={Campus Universit\'ario Darcy Ribeiro}, 
		city={Brasília},
		postcode={70910-900}, 
		state={Federal District},
		country={Brazil}}
	
	\affiliation[BSC]{organization={Barcelona Supercomputing Center},%Department and Organization
		addressline={Pla\c{c}a d'Eusebi G\"uell, 1-3, Les Corts, 08034}, 
		postcode={93022-750}, 
		city={Barcelona},
		% state={Rio Grande do Sul},
		country={Spain}}
	\affiliation[VT]{organization={Commonwealth Cyber Initiative},%Department and Organization
		addressline={Virginia Tech, 900 N. Glebe Rd. Arlington}, 
		% city={Virginia},
		postcode={22203}, 
		state={Virginia},
		country={United States of America}}

	\begin{abstract}
	Several machine learning-based Network Intrusion Detection Systems (NIDS) have been proposed in recent years. Still, most of them were developed and evaluated under the assumption that the training context is similar to the test context. This assumption is false in real networks, given the emergence of new attacks and variants of known attacks. To deal with this reality, the open set recognition field, which is the most general task of recognizing classes not seen during training in any domain, began to gain importance in machine learning based NIDS research. Yet, existing solutions are often \blue{bound} to high temporal complexities and performance bottlenecks. In this work, we propose an algorithm to be used in NIDS that performs open set recognition. Our proposal is an adaptation of the single-class Energy-based Flow Classifier (EFC), which proved to be an algorithm with strong generalization capability and low computational cost. The new version of EFC correctly classifies not only known attacks, but also unknown ones, and differs from other proposals from the literature by presenting a single layer with low temporal complexity. Our proposal was evaluated against well-established multi-class algorithms and as an open set classifier. It proved to be an accurate classifier in both evaluations, similar to the state of the art. As a conclusion of our work, we consider EFC a promising algorithm to be used in NIDS for its high performance and applicability in real networks. 
    
    \end{abstract}

	\begin{highlights}
		\item New Machine Learning Algorithm: Multi-class Energy-based Flow Classifier (EFC)
		\item New ensemble design transforms the single-class EFC into a multi-class algorithm
		\item Multi-class EFC performance was compared to other classifiers in closed-set datasets
		\item Multi-class EFC excelled in identifying unknown attacks in open-set datasets
	\end{highlights}
	
	
	\maketitle
	\section{Introduction}
	
	As organizations and individuals become increasingly connected, network attacks become more dangerous for victims and more attractive to cybercriminals. Security reports from the early 2020s, including the ENISA Threat Landscape \cite{ENISA}, DCMS's Cyber Security Breaches Survey \cite{DCMS}, and the ACSC Annual Cyber Threat Report \cite{ACSC}, have demonstrated the devastating impact of cyber attacks ranging from the exposure of personally identifiable information to extortion of millions of dollars in ransom payments. With the sophistication of threat capabilities increasing as we move deeper into the digital age, developing better security systems is critical.
    
    Currently, an important component of security systems is \ac {NIDS}. A \ac{NIDS} is software used in conjunction with firewalls and antimalware systems to protect networked devices from several threats \cite{Tidjon2019}. They can be implemented for different purposes, such as detecting network anomalies or detecting and classifying them. This latter descriptive approach is particularly interesting because categorizing intrusions enables the formulation of effective incident response actions, increasing the overall system performance~\cite{AHMAD2021102122}. 
    
    Although NIDS research has advanced a lot in recent years, some issues with state-of-the-art systems are still present in the literature. As argued by Zhang \textit{et al.} \cite{Zhang2021} and Apruzzese \textit{et al.} \cite{apruzzese2022cross}, the majority of papers in NIDS evaluate their proposals under closed-set experiments, \ie using the same network context in both train and test. This type of assessment does not reflect real networks, as it ignores the presence of new types of attacks or variations of known attacks. To develop an effective NIDS, this problem must be taken into account. 
    
    Some descriptive \ac{NIDS} capable of identifying unknown traffic have been proposed, \eg \cite{Zhang2021}, \cite{Buczak2016}, and \cite{pontes2019new}. These systems can recognize samples of classes that were not seen in the training stage, which is called open set recognition. In particular, Zhang \textit{et al.} \cite{Zhang2021} have shown great results using neural networks to perform open set classification in the intrusion detection domain. However, these proposals are often bound to high temporal complexities and performance bottlenecks due to the techniques they employ, like deep learning algorithms or cascading supervised classifiers. Naturally, this is not desirable since fast traffic analysis is a key requirement for \ac{NIDS} \cite{Buczak2016}. Considering these limitations, a \ac{NIDS} capable of detecting unknown attacks and classifying intrusions with low temporal and computational complexity is needed. 
    
    We propose a novel classifier to detect unknown attacks and classify known intrusions. Our model consists of an adaptation of the single-class \ac{EFC} \cite{pontes2019new}, which proved to be a method with good generalization capacity and low computational cost. Also, the EFC single-class classifier exhibits inherent open-set recognition capabilities, as discussed in \cite{pontes2019new}. The algorithm we present inherits the advantages of the original method while extending its functionality to the identification and categorization of several attack types. We performed experiments comparing the EFC with classical machine learning multi-classifiers and with current proposals from the literature, both in closed-set and open-set experiments. 
    
    The development of the Multi-class EFC aimed to explore the potential for detecting previously unknown or unclassified attacks. To assess its effectiveness, we compared the Multi-class EFC's performance against publicly available results reported by the authors of OCN \cite{Zhang2021} and ODIN \cite{Liang2018}, considering the same dataset used CICIDS2017. Also, considering the superiority of OCN compared to ODIN, we tested OCN against our solution in a new dataset, the CICIOMT2024~\cite{CICIoMT2024} in the context of \ac{IoMT}, considering multi-protocol attack vectors \blue{affecting} device security, exposing the potential of the multi-class EFC in this context. The main contributions of our work are:
	
	\begin{itemize}
		\item The proposal and development of the multi-class Energy-based Flow Classifier;
		\item A performance comparison between EFC and other multi-classifiers in closed-set experiments;
		\item An assessment of different multi-class classifiers' ability to correctly identify unknown attacks in open-set experiments;
		\item A performance comparison between the Multi-class EFC with current solutions from the literature.
	\end{itemize}
	
	The remainder of this paper is organized as follows: section \ref{sec:relatedwork} briefly summarizes some recent work on open-set \ac{NIDS}. Section \ref{sec:efc} presents the fundamentals of the statistical framework employed by EFC and the implementation of the multi-class EFC. Section \ref{sec:methodology} describes the methodology adopted in this work, including experiments, evaluation metrics, and dataset. Section \ref{sec:results} presents and discusses the results of the experiments. Finally, section \ref{sec:conclusion} closes the paper with conclusions and suggestions for future research.
	\section{Related work}
	\label{sec:relatedwork}
	
\begin{figure}[!ht]
		\centering
		%\includesvg[width=\columnwidth]{fig1.pdf}
		\includegraphics[width=0.9\columnwidth]{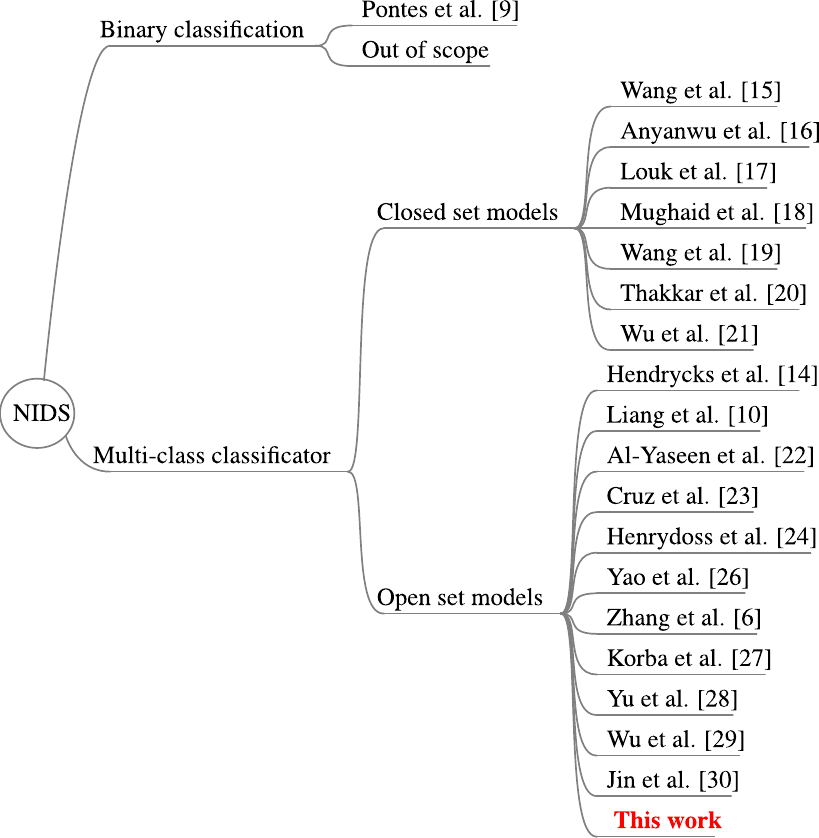}	
		\caption{NIDS Literature Review}
		\label{fig:taxonomy}
	\end{figure}

	In this section, we present an overview of the literature related to our work. In recent years, a wide variety of \ac{NIDS} was developed using \ac{ML} classifiers. We review some of these works focusing on descriptive classifiers, \ie which perform multi-classification, and on open set classifiers, which can detect unknown attacks. To coordinate the discussion and to position our work, we review the literature about \ac{NIDS} considering the proposed taxonomy presented in Figure \ref{fig:taxonomy}. 
    
    Binary \acp{NIDS} classify traffic into two classes: benign or malicious. They are useful to identify and block intrusions, but they do not provide information about identified attacks. We highlight \ac{EFC} as it was first presented by Pontes \etal \cite{pontes2019new}, as a binary classifier. The performance of EFC was evaluated considering the CICIDS2017, CIDDS-001 and CICDDoS2019 datasets and it was compared with several other \ac{ML} classifiers. From these experiments, the authors concluded that \ac{EFC} is capable of detecting anomalies in the three datasets with low computational cost and is also a robust algorithm for distribution changes. Due to these characteristics, we propose to modify the original single-class EFC method to perform multi-class classification by using the Potts model \cite{Wu1982} to model not only benign flows but also several attack classes.
    
    In addition to the feature of multi-classification, we also implemented a mechanism to identify unknown attacks. The ability to detect unknown classes is critical for modern network intrusion detection systems and is a particular case of a problem called open set recognition, which is the more general task of recognizing classes unseen in training in any domain~\cite{openset}. The open set field was initially developed for computer vision applications where it has been an important research topic with several notable works. 
    
    Hendrycks \textit{et al.} \cite{Hendrycks2017} proposed the Baseline method that uses probabilities from softmax distributions to identify unknown samples. They assessed the Baseline performance in tasks in computer vision, natural language processing, and automatic speech recognition, showing the effectiveness of this baseline across all these domains. Later, the ODIN \cite{Liang2018} method was proposed by Liang \textit{et al.} establishing a new state-of-the-art performance on the open set image recognition task. ODIN reduces the false positive rate compared to the Baseline from 34.7\% to 4.3\% on the DenseNet when the true positive rate is 95\%. 
	
	Meanwhile, in the field of intrusion detection, open-set recognition hasn't been widely explored. On the contrary, most experimental evaluations of machine learning-based NIDS are done in closed-set experiments \cite{apruzzese2022cross} and, under this assumption, traditional ML classifiers like \ac{SVM} \cite{Wang2023svm}~\cite{Anyanwu2023}, \ac{DT} \cite{Louk2023}~\cite{Mughaid2023} and \ac{ANN}~\cite{Wang2023} \cite{Thakkar2023} \cite{Wu2023} have already achieved almost perfect rates. Despite of that, in recent years, some works on NIDS have tried to develop more realistic models with more robust assessments, \ie exploiting the open set field. 
    
    Al-Yaseen \textit{et al.} \cite{Al-Yaseen2017} proposed a multi-level \ac{IDS} composed of five cascading layers, each identifying an attack class using \ac{SVM} or \ac{ELM} classifiers. After filtering all classes, the final layer classifies the remaining samples as \textit{Normal} or \textit{Unknown} with a \ac{SVM} classifier. The model overall detection rate in NSL-KDD dataset was 95.17\%. In our work, we also classify intrusions into known classes and a \textit{Suspicious} class, but the complexity of our algorithm is considerably lower than \blue{theirs}, especially in the classification phase. 
    
    Cruz \textit{et al.} \cite{Cruz2017} applied the Weibull-calibrated SVM (W-SVM) classifier on the KDDCUP’99 dataset in an open set experiment. They performed a comparison between a multi-class closed set Gaussian RBF kernel Platt-calibrated SVM, and a multi-class open set Gaussian RBF kernel W-SVM. They concluded that the accuracy of the two models is similar, but by weighting the accuracy to give more importance to unknown samples, the W-SVM exhibits better results. 
    
    On the basis of \cite{Cruz2017}, Henrydoss \textit{et al.} \cite{Henrydoss2017}
	investigated the EVM classifier in the intrusion detection domain. EVM is a generic multi-class classifier theoretically derived from the statistical Extreme Value Theory and developed in the context of computer vision \cite{Rudd2018}. Henrydoss \textit{et al.} applied EVM to the KDDCUP’99 dataset, testing it on different degrees of openness\blue{,} \ie on sets with different proportions of unknown data. They compared its accuracy with the W-SVM \cite{Cruz2017} and concluded that they have similar performances in every degree of openness. 
    
    Yao \textit{et al.} \cite{Yao2019} developed a novel \ac{IDS} framework based on \ac{HMLD}, evaluated on the KDDCUP’99 dataset. Their model classification phase consists of filtering each attack with a specific classifier trained to detect this attack. The chosen classifiers were \ac{SVM} and \ac{ANN}. Samples not classified as belonging to any of the known classes are called \textit{Impurity Data}. A small subset of the \textit{Impurity Data} is provided to a specialist to be labeled in new attack classes. Afterwards, this labeled set is used to train a \ac{DT} to classify the \textit{Impurity Data}. The overall model accuracy in the KDDCUP’99 dataset was 96.70\%. In our work, we also create a separate class for unknown intrusions, but our process is completely automatic, not requiring the \blue{labeling} by a specialist. In addition, we also achieve a lower temporal complexity than \ac{ANN} and \ac{SVM}. 
    
    More recently, Zhang \textit{et al.}  \cite{Zhang2021} proposed what we considered to be the \blue{state-of-the-art} in open set intrusion detection algorithms. The Open set Classification Network (OCD) is an \ac{ANN} based on the convolutional neural network that adopts the nearest class mean (NCM) classifier and uses both fisher loss and MMD loss to jointly optimize the CNN model. In addition, they proposed a method to discover new attacks among the unknown attack instances detected by OCN and to \blue{incrementally} learn 
	the classification of the discovered attack. They conducted several experiments in the KDDCUP’99 dataset and CICIDS2017 dataset, showing that their classifier \blue{outperforms} the \blue{state-of-the-art} detectors. In our work, we compared the performance of the EFC with the OCN in terms of detecting unknown attacks in the CICIDS2017 dataset. 
    
    The research by Korba \textit{et al.}  \cite{korba} proposes an unsupervised privacy-preserving FL with a Deep Auto-Encoder (DAE) for unknown activity, using open-set federated learning to classify the attacks with \blue{a} deep multi-class data descriptor. In the realm of IoV, it is the first, applying a \blue{new} training scheme with blockchain technology based on the Byzantine Fault Tolerance mechanism named \blue{Proof-of-Accuracy (PoA)}. All assessments were conducted on the 5G-NIDD and VDoS datasets for 0-day and N-day attack detection. However, the authors have not made the source code available, making a direct comparison with our work impossible. This is unfortunate, as they also utilized CICFlowMeter, which would have provided valuable insights for comparison. 
    
    The work proposed by Yu \textit{et al.}  \cite{DeepQN} employs a reconstruction error for open-set recognition, which is modeled as a discrete-time Markov decision process in Industrial Internet of Things (IIoT) environments. Techniques like target networks, double Q-learning, and experience replay are employed to improve performance in classifying known traffic and recognizing unknown attacks. Additionally, a Conditional Variational Autoencoder (CVAE) is integrated into the DQN's value network to handle both fine-grained traffic classification and unknown attack detection. Unfortunately\blue{,} the authors have not made the source code publicly available, which limits the ability to perform a comprehensive comparison. 
    
    The article by Wu \textit{et al.} \cite{Dandelion} proposes the Open-Set Dandelion Network (OSDN), which uses unsupervised heterogeneous domain adaptation to transfer intrusion detection knowledge from a data-rich source domain to the data-scarce IoT domain. The OSDN model can detect both known and newly-emerging intrusions not present in the source domain. Outperforming state-of-the-art methods by 16.9\%, it uses a "dandelion-like" feature space where each intrusion category is compactly grouped, and different categories are separated to ensure intra-category compactness and inter-category separability. Since the proposed dandelion framework utilizes the CICIDS2017 dataset, employing additional databases for testing and comparison could offer new avenues for evaluation in our research. However, the authors have unfortunately not made the source code available for use\blue{,} and we were unable to reproduce their work for comparison purposes. 
    
    Jin \textit{et al.} \cite{Jin2023} addresses the open set recognition problem by proposing an evolution cycle for \blue{IDSs} (EIDS) that incrementally incorporates newfound attacks into known attacks. The proposal is based on the federated learning paradigm \cite{Rahman2020}, which \blue{consists of} sharing data among different detectors distributed over the network to make each detector knowledge more robust. In their experiments, they employed three servers as distributed detectors and used the discriminative auto-encoder as a classification algorithm. The experimental results on ToN-IoT, BoT-IoT, and NSL-KDD datasets show that the overall model accuracy and the recall of unknown attacks were above 0.89 and 0.69 in all three datasets, respectively. 
    
    \blue{In the context of datasets to evaluate open set solutions, the CICIDS2017 dataset, created by the University of New Brunswick in 2017 by Sharafaldin \textit{et al.}~\cite{Sharafaldin2018}, provides simulated network traffic that includes a variety of up-to-date attacks alongside benign traffic, designed to emulate real-world network environments.
This dataset is available in both original PCAP files and CSV files, the latter containing 80 features extracted by CICFlowMeter, which encompass flow header information and empirical attributes from the network traffic.} \blue{Dadkhah \textit{et al.} \cite{CICIoMT2024}}, in turn, introduced a comprehensive benchmark dataset, CICIOMT2024, specifically designed for assessing the security of \ac{IoMT} devices. The authors detail the methodology used to create the dataset, which involved capturing network traffic from a testbed of 40 real and simulated IoMT devices operating over Wi-Fi, MQTT, and Bluetooth protocols. This captured traffic includes both benign activity and data from 18 distinct cyberattacks deliberately executed against the devices, categorized into five major classes: DDoS, DoS, Recon, MQTT, and spoofing. A primary contribution of this work is the \blue{provision of a} realistic multi-protocol dataset that reflects diverse attack vectors and device behaviors relevant to modern healthcare environments. Consequently, CICIOMT2024 serves as a resource to be tested in open-set experiments. Therefore, CICIOMT2024 is one of the datasets considered in this work for \blue{evaluation}. 
    
    Although in its early days, there has been some development in open set classifiers in the area of intrusion detection. However, to the best of our knowledge, all of these proposals are complex algorithms, such as cascade systems with several classifiers, like the Multi-level Hybrid model from \cite{Al-Yaseen2017} and the HMLD framework from \cite{Yao2019}; or deep neural network methods, like Baseline \cite{Hendrycks2017}, ODIN \cite{Liang2018}, OCN \cite{Zhang2021} \blue{and EIDS} with discriminative autoencoder \cite{Jin2023}; or SVM-based methods, like W-SVM \cite{Cruz2017} and EVM \cite{Henrydoss2017} with high processing complexity. All these proposals are bound to performance problems and may not be useful outside the research field, 
	as part of real-time NIDS. We propose a new classifier that stands out from other open-set proposals for having a simple and effective classification mechanism for both known and unknown classes with low temporal complexity. In the following section, we present the conceptual foundations of \ac{EFC}  \cite{pontes2019new} and the modifications made for the multi-class version.
	\section{Energy-based flow classification}
	\label{sec:efc}
	
	In this section, we present the energy-based classification technique. First, in subsection \ref{sec:efc_model_inference}, we briefly explain the EFC method as it was originally developed in \cite{pontes2019new}. Although a succinct explanation of the EFC method, it nonetheless gives the reader a reasonable understanding of it, providing some key definitions of the method, which will be necessary to understand the proposal of this work. Afterward, in subsection \ref{sec:efc_problem_definition}, we present the adaptation of the former method to perform multi-class classification and open-set detection.
	\subsection{Model inference}
	\label{sec:efc_model_inference}
	
	The main idea of \ac{EFC}'s training phase is to perform a mean field variational (Bayesian) inference to find the posterior probability distribution underlying the flow class to be detected. Once the posterior distribution is defined, it is used to classify new flows. This is done by calculating a quantity called \textit{flow energy}, which is a measure of how unlikely a flow is to belong to a given probability distribution. The definition of \textit{energy} was kept unchanged as a quantity coming from the original inference problem from statistical physics (the Potts model), which was concerned with atomic spins in a lattice and served as inspiration for the development of the EFC. Thus, to define and explain this measure, we will present the inference process of a generic flow class distribution. 
    
    Let $k = (a_{1}, …, a_{n})$ be a network flow, where each position $i \in \node = \{1, ..., n\}$ represents a feature and each feature can assume values $a_{i} \in \Omega = \{1, ..., Q\}$. As a practical example, a network flow can be considered as any set of \blue{packets} that, by some convention, share the same characteristics or simply features, such as source IP, incoming port, or payload size. Let $\flowidset$ be the set of all possible flows from a given class and $\flowset \subset \flowidset$ the subset from which we want to infer the distribution. The probabilistic model that best represents $\flowidset$ is the one that, respecting empirical observations of $\flowset$, assumes as little prior information as possible. Equivalently, it is the one that least restricts uncertainty among all possible models. Thus, using entropy as a measure of uncertainty, we want to find the distribution that maximizes the entropy while respecting the observed characteristics of $\flowset$. Formally, we want to solve the following problem of maximizing entropy
	
	\begin{align}
		\max_{P} & \ \ \ - \sum_{\flowid \in \flowidset}  \ \ \ \ \ \ \  P(\feature_{\flowid1}...\feature_{\flowid \featureids}) log (P(\feature_{\flowid 1}...\feature_{\flowid \featureids})) \label{eq:max}\\
		s.t.\nonumber\\
		& \ \ \ \sum_{\flowid \in \flowidset | \feature_{\flowid \featurei} = \feature_{\featurei}}  \ \ \ \ P(\feature_{\flowid 1} ... \feature_{\flowid \featureids}) = f_\featurei(\feature_\featurei) \label{eq:3} \\ 
		&\qquad\qquad\forall \featurei \in \node;\ \forall \feature_{\featurei} \in \Omega; \nonumber \\
		&\sum_{\flowid \in \flowidset | \feature_{\flowid \featurei} = \feature_{\featurei}, \feature_{\flowid \featurej} = \feature_{\featurej}} P(\feature_{\flowid 1} ... \feature_{\flowid \featureids}) = f_{\featurei \featurej} (\feature_\featurei, \feature_{\featurej}) \label{eq:4} \\
		&\qquad\qquad\forall (\featurei,\featurej) \in \node^2\ |\ \featurei \neq \featurej;\ \forall (\feature_{\featurei}, \feature_{\featurej}) \in \Omega^2;\nonumber
	\end{align}
	
	where $f_\featurei(\feature_\featurei)$ is the empirical frequency of value $\feature_\featurei$ on feature $\featurei$ and $f_{\featurei\featurej}(\feature_\featurei,\feature_{\featurej})$ is the empirical joint frequency of the pair of values $(\feature_\featurei,\feature_{\featurej})$ of features $\featurei$ and $\featurej$ \blue{observed in $\flowset$}. In other words, we seek the distribution of greater entropy that reflects the configuration of flow features in $\flowset$. The proposed maximization can be solved using a Lagrangian function such as presented in \cite{Jaynes1957}, yielding the following Boltzmann-like distribution: 
	\begin{equation}
		P^*(\feature_{\flowid1}...\feature_{\flowid\featureids}) = \frac{e^{- \mathcal{H}(\feature_{\flowid1}...\feature_{\flowid\featureids})}}{Z} \label{eq:opt_dist}
	\end{equation}
	where
	\begin{equation}
		\mathcal{H}(\feature_{\flowid1}...\feature_{\flowid\featureids}) = - \sum_{\featurei,\featurej \mid \featurei<\featurej} \coup_{\featurei\featurej}(\feature_{\flowid\featurei},\feature_{\flowid\featurej}) - \sum_\featurei \field_\featurei(\feature_{\flowid\featurei}) \label{eq:hamil}
	\end{equation}
	is the Hamiltonian of flow k and Z is the partition function that normalizes the distribution. We will ignore $Z$ as we are not interested in calculating specific flow probabilities. In fact, we are only interested in the Hamiltonian of a flow, which is exactly the measure we call \textit{energy}. Before discussing its definition and the functions $ \coup_{\featurei \featurej}(\cdot)$ and $\field_\featurei(\cdot)$, note that there is an important relationship between the energy of a flow and its probability, given by Equation \ref{eq:opt_dist}: the higher the flow energy, the lower its probability. This relationship implies that the energy of a flow is a measure of how unlikely it is to belong to that distribution. So, if we infer the distribution for a flow class and calculate the energy of a new flow with respect to this distribution, we get a measure of how likely it is that it belongs to that class. For this reason, the energy allows us to classify a given flow as belonging to a certain class or not. 
    
    Note that, by the solution presented in \cite{Jaynes1957}, the energy is completely defined by the Lagrange multipliers $ \coup_{\featurei \featurej}(\cdot)$ and $\field_\featurei(\cdot)$, associated \blue{with} constraints (\ref{eq:3}) and (\ref{eq:4}). So, to infer the distribution and to be able to compute energy values, we need to calculate $\field_\featurei(\cdot)$ and $\coup_{\featurei \featurej}(\cdot)$, defined in \cite{Jaynes1957} as
	
	\begin{align}
		\field_\featurei(\feature_\featurei) = ln \left (\frac{f_\featurei(\feature_\featurei)}{f_\featurei(Q)} \right ) - \sum_{\featurej, \feature_\featurej}\coup_{\featurei\featurej}(\feature_\featurei, \feature_\featurej)f_\featurej(\feature_\featurej) 
		\label{eq:mf2}
	\end{align}
	and
	\begin{align}
		\coup_{\featurei \featurej}(\feature_\featurei,\feature_{\featurej}) = -(C^{-1})_{\featurei \featurej} (\feature_\featurei,\feature_{\featurej}),\label{eq:coupling}\\ \forall (\featurei, \featurej) \in \node^2, \forall (\feature_\featurei, \feature_\featurej) \in \alphabet^2, \feature_\featurei, \feature_\featurej \neq Q \nonumber
	\end{align}
	where
	\begin{equation}
		C_{\featurei \featurej}(\feature_\featurei,\feature_{\featurej}) = f_{\featurei\featurej}(\feature_\featurei,\feature_{\featurej}) - f_\featurei(\feature_\featurei)f_{\featurej}(\feature_{\featurej})
	\end{equation}
	is the covariance matrix obtained from single and joint empirical frequencies. 
    
    In an intuitive way, let $\field_\featurei(\cdot)$ and $\coup_{\featurei \featurej}(\cdot)$ be defined for a subset $\flowset$ according to equations~\ref{eq:mf2} and \ref{eq:coupling}. Let $k = (a_{1}, …, a_{n})$ be a flow, where each position $ i \in \{1, ..., n \}$ represents a feature and each feature can assume values $ a_ {i} \in \{1, ..., Q \}$. Then, the local fields $\field_\featurei(\feature_\featurei)$ are a measure of how likely it is that feature $i$ assumes the value $a_ {i}$ in $\flowset$. Similarly, for the same flow, the coupling values $\coup_{\featurei \featurej}(\feature_\featurei,\feature_{\featurej})$ are a measure of how likely it is that features $i$ and $j$ assume, at the same time, the values $a_i$ and $a_j$ in the set $\flowset$. Therefore, the sum of \blue{coupling} values and local fields of all features of \textit{k} reflects the similarity of the flow with the original subset $\flowset$ feature by feature. 
    
    To train EFC's model, a set of latent variables must be inferred. This set of latent variables defines a posterior distribution, which is a probability distribution conditioned to the observed (training) data. Basically, two sets of variables need to be inferred for each flow class: 1st-order variables, or local fields $\field_\featurei $, and 2nd-order variables, or coupling values $\coup_{\featurei \featurej}$. 
    
    \blue{Further}, in the classification phase, it is possible to calculate the energy of a new flow using the previously inferred latent variables for a given flow class in the optimality using Equation (\ref{eq:hamil}). The energy of a flow $k$ is a linear combination of local fields and coupling values for all features and features pairs of flow $k$, and it reflects how likely it is that this exact configuration of features values occurs in the flow set $\flowset$. If the flow energy is high, it means that it has a low probability of belonging to that class -- in other words, it does not resemble the flows that generated the posterior distribution for that class. Likewise, if the energy is low, the flow is more likely to belong to the set of flows that generated the posterior distribution for that class. 
    
    To decide whether the energy is high or low, we use a threshold defined by the 95th percentile of the energies of the samples used to infer the model. In single-class \ac{EFC}, the training is done with benign samples only. Therefore, the classification is performed with respect to the normal distribution: if the flow has lowest energy than the 95th percentile of the benign samples, \ie below the threshold, it is considered to be normal. Otherwise, it is labeled as abnormal traffic.
	
	For more details on the development of single-class \ac{EFC}, as well as a complete explanation of the model inference, please refer to \cite{pontes2019new}. Next, we will present our proposed multi-class version of \ac{EFC} and an algorithm for its implementation. 
    
	\subsection{Problem definition}
	\label{sec:efc_problem_definition}
	
	\newcommand{\class}{k}
	\newcommand{\classSet}{\mathcal{K}}
	\newcommand{\classMax}{K}
	\newcommand{\threshold}{\tau}
	\newcommand{\flowT}{f}
	\newcommand{\flowTSet}{\mathcal{F}}
	\newcommand{\flowTSetMax}{F}
	\newcommand{\flowTAux}{g}
	\newcommand{\hamil}{\mathcal{H}}
	\newcommand{\varSelectClass}{x}
	\newcommand{\varSuspiciousFlow}{y}
	\newcommand{\largeNumber}{\Psi}
	
	Consider a set of possible classification classes $\class \in \classSet =\{1,\cdots,\classMax\}$, such that $\classMax$ represents the total number of classes available for a given problem. Let $\flowTSet_{\class}$  be the training set of all flows labeled $\class$ such that $\flowT \subset \flowTSet_\class = \{1, ..., \flowTSetMax\}$. For all $\flowT \in \{1, ..., \flowTSetMax\}$, we infer the coupling values $\coup^\flowT_{\featurei \featurej}$ and local fields $\field^\flowT_{\featurei}$ from $\flowTSet_{\class}$. We compute the energy vector $\hamil_{\class 1}, ..., \hamil_{\class\flowTSetMax}$, where $\mathcal{H}_{\class\flowT}$ is the Hamiltonian from Equation (\ref{eq:hamil}) for a given flow $\flowT$ in class $\class$. We also define the threshold $\threshold_{\class}$ as the 95th percentile of the energies of the samples in $\flowTSet_{\class}$ for a given class $\class$, sorted according to their respective energy. 
    
    Taking into account the model for each class $\class$ created during the EFC training and \blue{its} calculated threshold, we must determine the best-fit class for a given flow $\flow$. Therefore, consider a binary decision variable:
	
	\begin{align}
		\varSelectClass_{\class\flow} = \left\{ \begin{aligned}&1; &\ if\ flow\ \flow\ belongs\ to\ class\ \class; \\& 0; & otherwise;\end{aligned}  \right.
	\end{align}
	
	Also, consider that a given flow $\flow$ may be suspicious and does not belong to any known class of the set $\classSet$. In this sense, consider a binary decision variable:
	\begin{align}
		\varSuspiciousFlow_{\flow} = \left\{ \begin{aligned}&1; &\ if\ flow\ \flow\ is\ suspicious; \\& 0; & otherwise;\end{aligned}  \right.
	\end{align}
	considering these decision variables, the problem of selecting a specific class $\class$ of a given flow $\flow$ is constrained to the following:
	
	\subsubsection{Each flow can only belong to a known class or become suspicious}
	When a flow $\flow$ is tested, it can only be assigned to a single class $\class$ of the set of known classes $\classSet$. Otherwise, it is labeled suspicious. \blue{This} means that summing all possible binary decisions $\varSelectClass_{\class\flow}$ combined \blue{with} $\varSuspiciousFlow$ must result in 1.
	\begin{align}
		& \sum_{\class=1}^{\classMax} \varSelectClass_{\class\flow} + \varSuspiciousFlow = 1;\label{opt_mustSelect}
	\end{align}
	
	\subsubsection{A class will only be chosen if similar to the selected flow}
	For a given flow $\flow$, it can only be assigned to a class $\class$ of the set of known classes $\classSet$, if they are similar. A flow $\flow$ is only similar to a class $\class$ when its energy $\hamil_{\class\flow}$ is \blue{less} than or equal to the 95th percentile $\threshold_{\class}$ of the energies from that class. \begin{align}
		& \varSelectClass_{\class}\hamil_{\class\flow}\le\threshold_{\class};\ \forall \class \in \classSet; \label{opt_threshold}
	\end{align}
	
	\subsubsection{Decision variables are binary}
	Each decision regarding the labeling of a given flow $\flow$ as class $\varSelectClass_{\class\flow}$ or suspicious $\varSuspiciousFlow_{\flow}$ is modeled as a "yes" or "no" \blue{question}, resulting in binary decisions constrained between 0 and 1.
	\begin{align}
		& 0 \le \varSelectClass_{\class\flow} \le 1; \ \forall \class \in \classSet;\label{opt_domainSel}\\
		& 0 \le \varSuspiciousFlow_{\flow} \le 1\label{opt_domainSus}; \end{align}
	
	\subsubsection{Flow Class Decision Problem}
	
	The problem formulated starts with the Objective Function (\ref{opt_obj}), \blue{whose left term determines} that a flow $\flow$ belongs to the class $\class \in \classSet$ with the smallest value of energy $\varSelectClass_{\class\flow}\hamil_{\class\flow}$, \ie the most similar class to which a flow belongs. However, given the Constraint (\ref{opt_threshold}), a flow may not be similar to any known class considered. Combined with the constraint (\ref{opt_mustSelect}), the decision variable $\varSuspiciousFlow_{\flow}$ for the suspicious flow becomes 1, since it is not possible to select any other similar class. Finally, in the right term of the objective, $\varSuspiciousFlow_{\flow}$ will be multiplied by a large number $\largeNumber$, which will classify the flow $\flow$ as suspicious, \blue{a classification that the optimization will also seek to avoid}. % and take the minimum value $\mathcal{H}_{r}$ of that vector. 
	\begin{align}
		\min_{\flow} &\ \ \ \  \sum^{\classMax}_{\class=1}\varSelectClass_{\class\flow}\hamil_{\class\flow} + \varSuspiciousFlow_{\flow}\largeNumber; \label{opt_obj} \\
		s.t. \nonumber\\
		&(\ref{opt_mustSelect}), (\ref{opt_threshold}), (\ref{opt_domainSel}, (\ref{opt_domainSus});\nonumber
	\end{align}
	
	% If $\mathcal{H}_{r} \le t_{r}$ we label the flow with class $r$. Otherwise, we label it as \textit{suspicious}. 
	
	To solve the problem defined in (\ref{opt_obj}) any integer linear solver, such as Google's OR-Tools\footnote{Google's OR-Tools: \url{https://developers.google.com/optimization}} or IBM's CPLEX\footnote{IBM's CPLEX: \url{https://www.ibm.com/products/ilog-cplex-optimization-studio}}, can be used. However, none of the considered solvers will present polynomial complexity. Also, they are packaged into large and general-purpose libraries, which can easily degrade \blue{software} performance for real-time processing use cases, such as line-rate network IDS. Finally, the solvers are not supposed to be incorporated into a simple library to be used in open experiments, which hinders their usage. Therefore, next, we propose a different approach to solve this problem and to implement the Multi-class EFC.
	\subsection{Multi-class EFC}
	\label{sec:proposal}
	
	The multi-class \ac{EFC} uses the same techniques as the single-class version, \ie the algorithm decides whether a flow belongs to a class by looking at its energy value. However, in the single-class version the flow energy is calculated only with respect to the normal distribution, resulting in a binary classification. In the multi-class case, several distributions are inferred, one for each flow class. Afterwards, flow energies are calculated in each distribution and their values are compared to return the classification result. \begin{figure}[h]
		\centering
		\includegraphics[width=0.8\columnwidth]{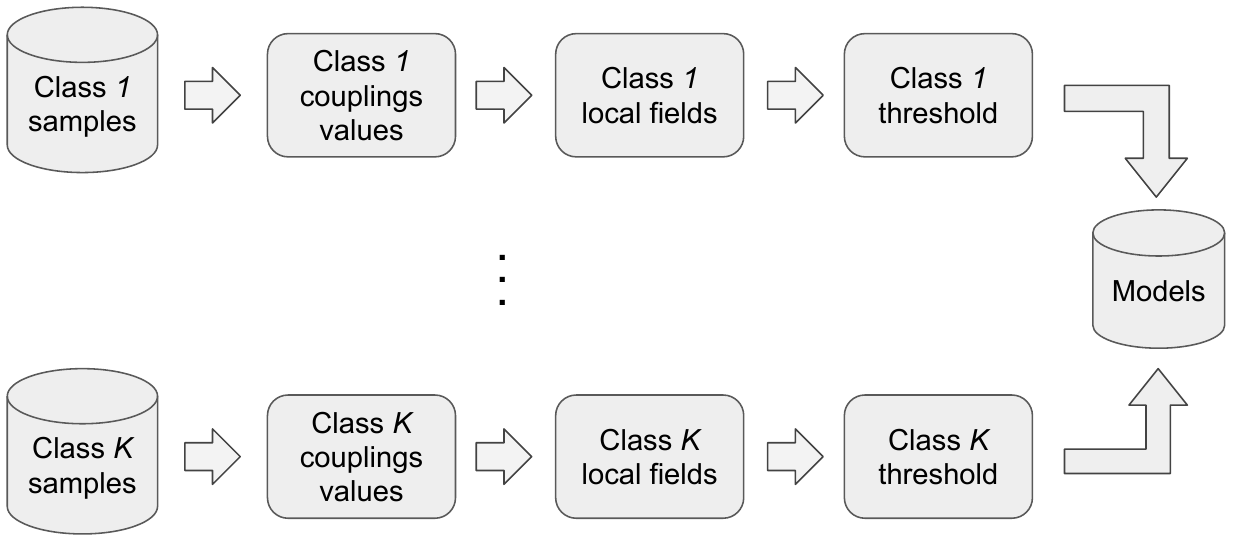}
		\caption{Multi-class EFC training phase}
		\label{fig:EFCtrain}
	\end{figure}

	Figure \ref{fig:EFCtrain} shows the multi-class \ac{EFC} training process. This phase consists of a replication of the single-class training process to more than one class. While in the single-class version we infer a model only for the benign traffic class, here we need to infer a model for each flow class. So, initially, training samples are grouped by class. Then, the models are inferred and the thresholds are computed for each class, in the same way as the single-class version, \ie calculating the local fields, coupling values and assuming a statistical threshold, namely the 95th percentile of training \blue{sample energies}. Lastly, the models induced for each class are stored to be used in the classification phase.
	\begin{figure}[!htb]
		\centering
		\includegraphics[width=0.8\columnwidth]{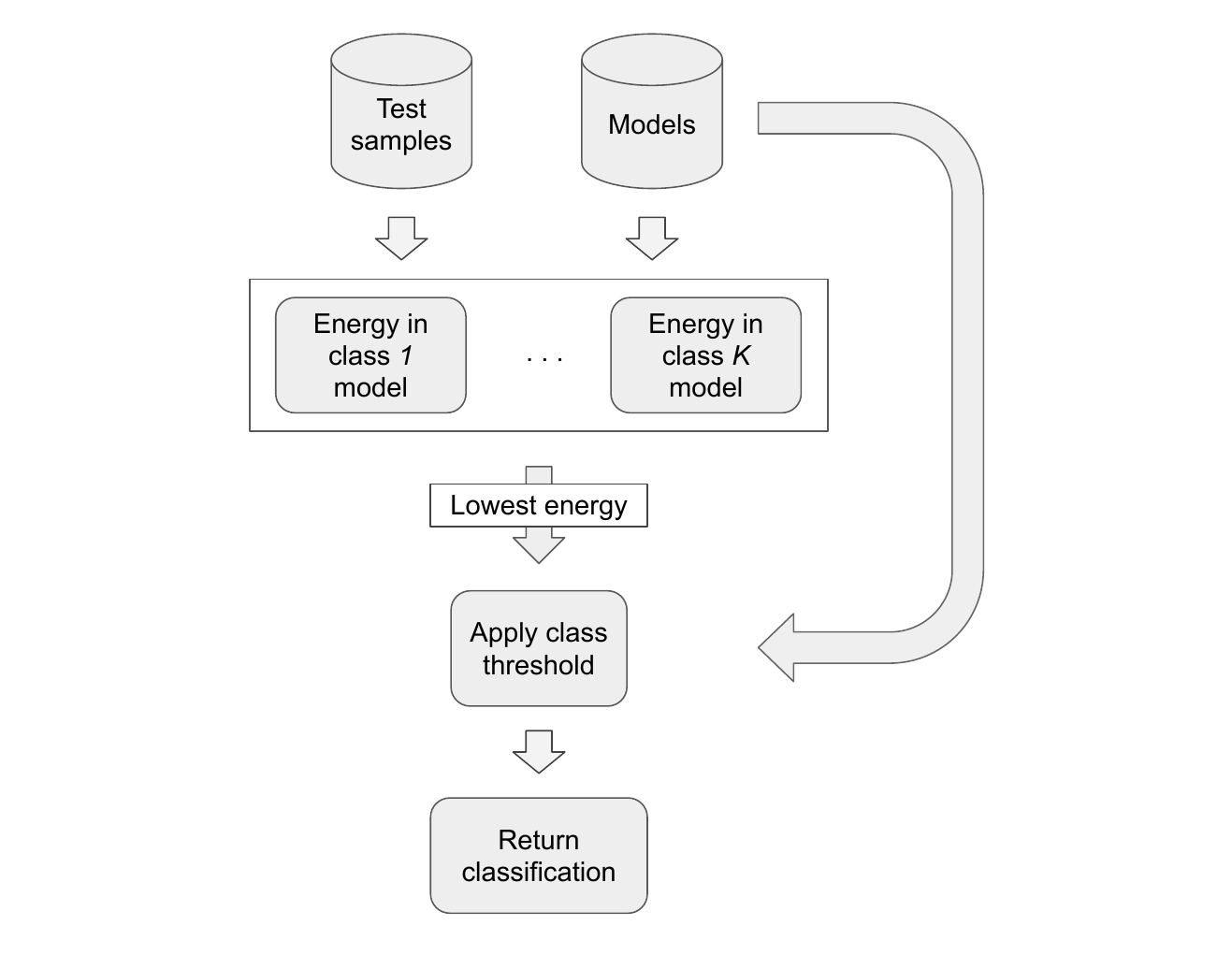}
		\caption{Multi-class EFC testing phase}
		\label{fig:EFCtest}
	\end{figure}
	
	Figure \ref{fig:EFCtest} shows the classification process of the multi-class \ac{EFC}. To classify an instance from the test set, its energy is computed in each model induced in the training phase, generating an energy vector for each instance. As explained in the previous subsection, the energy of a flow in a distribution is a measure of the dissimilarity of that flow to the set used to infer the distribution. So, the energy vector of a flow actually contains values inversely proportional to the probabilities of the flow belonging to each class. Therefore, after computing the energies, EFC takes the lowest generated value and compares it with the threshold of that class (since \blue{lower} energy corresponds to \blue{higher} similarity). If the energy is below the threshold, the flow is considered to be from the class that generated the energy, otherwise, it is classified as \textit{suspicious}. This second situation means that even the class that most closely resembles the flow isn't similar enough to it. Therefore, the flow is considered to be suspicious, possibly corresponding to an unknown type of attack.
	\subsection{EFC Multi-Class Algorithm}
	
	\begin{algorithm}[!h]
		\hspace*{\algorithmicindent}
		\scalebox{0.8}{
			\begin{minipage}{\textwidth}
				\textbf{Input:} $flows_{(K \times N)}$, $labels_{(L)}$, $Q$, $\alpha$
				\begin{algorithmic}[1]
					\State import all model inference functions
					\For{$class$ in $labels$}
					\State $flow\_class$ $\leftarrow$ flows labeled with $class$
					\State $f\_i \leftarrow SiteFreq(flow\_class, Q, \alpha)$
					\State $f\_ij \leftarrow PairFreq(flow\_class, f\_i, Q, \alpha)$
					\State $e\_ij \leftarrow Couplings(f\_i, f\_ij, Q)$
					\State $h\_i \leftarrow LocalFields(e\_ij, f\_i, Q)$
					\State $cutoff \leftarrow DefineCutoff(flow\_class, e\_ij, h\_i, Q)$
					\State $Couplings.Add(e\_ij)$
					\State $LocalFields.Add(h\_i)$
					\State $Cutoffs.Add(cutoff)$
					\EndFor
					\While{Scanning the Network}
					\State $flow \leftarrow$ wait\_for\_incoming\_flow()
					\State $energies \leftarrow$ []
					\For{$class$ in $labels$}
					\State e $\leftarrow 0$
					\For{$i$ $\leftarrow 1$ to $N-1$}
					\State $a\_i \leftarrow flow[i]$
					\For{$j$ $\leftarrow i + 1$ to $N$}
					\State $a\_j \leftarrow flow[j]$
					\If{$a\_i \neq Q$ and $a\_j \neq Q$}
					\State $e \leftarrow e - Couplings[class][i,a\_i,j,a\_j]$
					\EndIf
					\EndFor
					\If{$a\_i \neq Q$}
					\State $e \leftarrow e - LocalFields[class][i,a\_i]$
					\EndIf
					\EndFor
					\State $energies.Add(e)$
					\EndFor
					\State $lowest \leftarrow min(energies)$
					\State $cutoff \leftarrow Cutoffs[energies.index(lowest)]$
					\If{$lowest \leq cutoff$}
					\State $label \leftarrow labels[energies.index(lowest)]$
					\Else
					\State $label \leftarrow unknown\  intrusion$
					\EndIf
					\EndWhile
				\end{algorithmic}
			\end{minipage}
		}
		\caption{Multi-class Energy-based Flow Classifier}
		\label{algo:efc}
	\end{algorithm}

	Algorithm \ref{algo:efc} shows the pseudo-code of the procedures described above. Lines 2 to 11 represent the EFC's training phase, in which the $\mathcal{F}_{s}$ sets are separated, the statistical models $\coup_{s_{\featurei \featurej}}$ and $\field_{s_{\featurei}}$ are induced and the threshold $t_s$ is defined, for each class $s \in \{1,...,l\}$. When a network flow is captured, lines 16 to 38 perform its classification. First, the energy vector $\mathcal{H}_{1}, ..., \mathcal{H}_{l}$ is computed, using each model inferred in the training phase. Then, in lines 28-32, select the lowest energy $\mathcal{H}_r$ and check if it is below the threshold $t_r$. If so, the flow is labeled as $r$. Otherwise, it is labeled as \textit{suspicious}. The classifier training complexity (lines 2-11) is $$O(L[M^3Q^3 + N M^2 Q^2])$$ where $N$ is the number of instances in the training set, $L$ is the number of classes, $M$ is the number of features \blue{and $Q$} is the size of the alphabet used for discretization, \ie the maximum number of bins obtained in the discretization. Considering that $N$ is expected to be much \blue{larger} than the number of classes, the number of features, and the size of the discretization alphabet, the term $L N M^2 Q^2$ is dominant over $L M^3 Q^3$\blue{,} and we can simplify EFC training complexity to $$O(L N M^2 Q^2)$$ Meanwhile, the complexity for \blue{the} classification phase (lines 16-38) is $$O(L M^{2}N)$$ Therefore, both training and testing complexities are linear in the number of samples and can scale up well. In the next section, we will discuss the datasets, metrics, and the experimental setup used to evaluate the classifier presented in this section.
	\section{Methodology}  
	\label{sec:methodology}
	
	This section describes in detail the methodology adopted in this work to evaluate the multi-class EFC. Subsection \ref{sub:exp} discusses the experiments carried out with the classifier, and subsection \ref{sub:data} presents the dataset in which the experiments were conducted.
	\subsection{Experiments}
	\label{sub:exp}
	
	We divided the evaluation of our solution into two parts. The first is a performance comparison of EFC against classical multi-class algorithms, such as DT, MLP, and SVM, in both closed-set and open-set experiments. The second is a reproduction of an open set experiment by Zhang \textit{et al.} \cite{Zhang2021} to compare EFC with the Baseline, ODIN, and OCN algorithms, all specially designed to detect unknown attacks. 
    
    To compare EFC with DT, SVM, and MLP, we performed a 5-fold cross-validation \blue{on} CICIDS2017 using their scikit-learn\footnote{Scikit learn library - https://scikit-learn.org} implementations with default hyperparameters. We also used EFC with its default hyperparameters: 30 for the number of discretization bins and 0.5 for pseudo-count weights. In the closed-set version of the experiment, we performed a traditional 5-fold cross-validation with the aim of evaluating EFC's ability to distinguish between classes of attacks as a simple multi-class classifier. The results of this assessment are shown in terms of F1 score, defined as the harmonic mean of Precision and Recall measures. 
    
    In the open set version of the experiment, we used the same sets from the previous experiment, but we systematically removed an attack class from the training sets, while keeping this class in the test sets. For example, for the \textit{DoS} class, we performed the 5-fold cross-validation having removed the \textit{DoS} samples from the training sets. In this way, the \textit{DoS} samples present in the test set became unknown attacks, as the classifier did not train with this attack. This procedure was executed for all classes of attacks present in the dataset, one at a time. Although the other algorithms do not have mechanisms to identify unknown attacks, they serve as baselines for what would be the behavior of most techniques encountered today. More details about data pre-processing for this experiment will be presented in the next subsection. 
    
    The second part of our evaluation consisted of reproducing the experiment by Zhang \textit{et al.} \cite{Zhang2021}. As in our open set experiment, they removed one attack class at a time from training, keeping that attack in the test set. However, only the attacks   \textit{DoS slowloris}, \textit{Bot}, \textit{Infiltration}, \textit{Web Attack}, \textit{DoS Slowhttptest} and \textit{Heartbleed} were considered in their experiment. The benign class and the other attacks were used as known data for all unknown attacks. They performed a simple train test split with 80\% of the known data for training and 20\% for testing, which was added with one unknown attack at a time. Despite their algorithm being multi-class, they binarized the labels, making the known data positive and the unknown attack data negative. They used the following threshold-independent metrics to evaluate their results: the area under the receiver operating characteristic curve (AUROC) and the area under the precision-recall curve (AUPRC). In the following subsection, we present the datasets used in these experiments and more details about the data preparation.
	
    \subsection{Datasets}
	\label{sub:data}
	
    This work utilizes two datasets: CICIDS2017 \cite{Sharafaldin2018} and CICIOMT2024 \cite{CICIoMT2024}. Both datasets were obtained from the Canadian Institute for Cybersecurity (CIC) repository at the University of New Brunswick\footnote{CIC UnB - https://www.unb.ca/cic/datasets/} and are described in detail below, along with their preparation.
	\subsection{CICIDS2017}
	CICIDS2017 \cite{Sharafaldin2018} is a dataset created by the University of New Brunswick in 2017. It contains simulated traffic in packet-based and bidirectional flow-based format. CICIDS2017 includes the most up-to-date attacks as well as benign traffic, providing an environment that resembles real networks. The dataset was made available in two formats: the original PCAP files and CSV files, which resulted from the extraction of 80 features from these \blue{PCAPs} by CICFlowMeter. These features consist of flow header information, such as \textit{Source IP}, \textit{Destination IP}, \textit{Source port}, \textit{Destination port} and \textit{Protocol}; and empirical attributes such as \textit{Duration}, \textit{Number of transmitted packets}, \textit{Number of transmitted bytes}, \textit{Flags}, and \textit{Date first seen}. It is worth mentioning that we considered all the available features returned by the CICFlowMeter\footnote{CICFlowMeter GitHub \blue{- }https://github.com/CanadianInstituteForCybersecurity/CICFlowMeter} tool without changes, \ie we considered 87 network features gathered, for example, protocol in use, average packet length, and flow duration. \begin{table}[!h]
		\caption{CICIDS2017 dataset composition in its original version and in the version used in \blue{Zhang} et al. \cite{Zhang2021}}
		\label{table:comp}
		\centering
		\input{tab1}
	\end{table}
	
	In the first experiment, comparing EFC with DT, MLP and SVM, we used the dataset with its original attributes and performed the following pre-processing steps. The features \textit{Flow ID, Source IP, Destination IP} and \textit{Time stamp} were removed, because they only make sense in the emulated environment and are not informative regarding the traffic nature. We also combined the classes \textit{Web Attack-Brute Force}, \textit{Web Attack-XSS} and \textit{Web Attack-Sql Injection} into one class called \textit{Web Attack}, as \blue{did Zhang et al.} \cite{Zhang2021}, because the behavior of flows of these classes is practically the same at the network level. After that, we encoded the labels and the symbolic features using ordinal encoding and normalized the continuous features by their maximum absolute value so that they fit in the range $[-1, 1]$. For EFC only, we added a final discretization step. The partitioning of the train and test sets was done to perform a 5-fold cross-validation in the whole dataset. Lastly, we undersampled the training sets to avoid data imbalance problems, restricting to 5.000 instances the classes that had more than this. 
    
    In the experiment from Zhang \etal\cite{Zhang2021}, we used their version of CICIDS2017\footnote{Zhang et al dataset - https://github.com/zhangzhao156/scalable-NIDS}, created by extracting 256-dimensional header features from the PCAP files based on the original feature extraction method. Zhang et al. \cite{Zhang2021} employed a different technique to process the results from the flow generation tool CICFlowMeter, resulting in variations in the number of flows and features compared to the original CICIDS2017 dataset, \blue{as presented} in Table \ref{table:comp}. As to the data preprocessing, we used the dataset exactly as it was released, just adding a discretization step since EFC requires discrete data. We partitioned the train and test sets exactly as described in \cite{Zhang2021}, with 80\% of the samples for training and 20\% for testing, and \blue{undersampled} the \textit{DDos}, \textit{Dos Hulk} and \textit{PortScan} classes to \blue{retain} 50.000 flows. All the scripts used to pre-process data and execute the experiments were made available in our project repository\footnote{EFC repository - https://github.com/EnergyBasedFlowClassifier/EFC}. Table \ref{table:comp} shows the composition of CICIDS2017 in both experiments. In the following section, we will present and discuss our results.
	\subsection{CICIOMT2024}
	\begin{table}[!h]
		\caption{CICIOMT2024 dataset composition in its original version Dadkhah et al.\cite{CICIoMT2024}}
		\label{table:compCICIOMT2024}
		\centering
		\input{tab2}
	\end{table}
	The CICIOMT2024\footnote{Dadkhah \etal provided by the Canadian Institute for Cybersecurity of the University of \blue{New} Brunswick  - https://www.unb.ca/cic/datasets/iomt-dataset-2024.html}~\cite{CICIoMT2024} dataset  was created by the Canadian Institute for Cybersecurity (CIC) at the University of New Brunswick, introduced in 2024. It provides a benchmark dataset focused on \ac{IoMT} security, containing network traffic captured from an IoMT testbed. This testbed included 40 devices (25 real and 15 simulated) and utilized multiple protocols common in healthcare, such as Wi-Fi, MQTT, and Bluetooth. 
    
    CICIOMT2024 includes data from 18 different cyberattacks -- categorized as DDoS, DoS, Recon, MQTT, and spoofing -- as well as benign traffic and device lifecycle profiling data (capturing power, idle, active, and interaction states), aiming to represent realistic IoMT environments. The dataset is available in its original packet capture (.pcap) format for both Bluetooth and Wi-Fi/MQTT traffic. Additionally, for the Wi-Fi/MQTT traffic, .csv files containing extracted features are provided, intended for use in \ac{ML} evaluations. These features include attributes such as Header-Length, Protocol Type, Duration, various flag counts (e.g., syn\_flag\_number, ack\_flag\_number), protocol types (e.g., TCP, UDP, MQTT), and various statistical measures derived from the traffic like IAT, \blue{Magnitude}, and Variance. The extracted features are made available to support the development and evaluation of security solutions using ML techniques. 
    
    To prepare CICIOMT2024 for experimentation, we conducted the same pre-processing used in CICIDS2017. We removed features from CICIOMT2024, such as \textit{Flow ID, Source IP, Destination IP} and \textit{Time stamp}. As can be seen in Table~\ref{table:compCICIOMT2024}, CICIOMT2024 presents a large number of samples that would take days \blue{or} months to execute model training for exponential complexity methods, such as \ac{SVM}. Therefore, we reduced the dataset using random sampling from all attack classes, taking a maximum of 5000 samples per class.
	\section{Results}
	\label{sec:results}
	
	Our assessments are two-fold: \1 a performance comparison with \ac{SVM}, \ac{DT} and \ac{MLP} in both closed-set and open-set experiments and \2 a performance comparison with Baseline, ODIN, and OCN in open-set experiments. In the following subsections, we present and discuss the results of both assessments.
	\subsection{Closed-set comparative analysis}
	\label{subsec:closed}
	\begin{table}[!h]
		\centering
		\input{tab3}
	\end{table}
	This experiment aims to characterize the \ac{EFC} as a multi-class method, comparing it with other consolidated multi-class classifiers. Table \ref{tab:compCICIDS2017} and Table \ref{tab:compCICIoMT2024} show the results of \ac{EFC}, \ac{DT}, \ac{MLP} and \ac{SVM} in 5-fold cross-validation on CICIDS2017 and CICIOMT2024 datasets, respectively. Each table shows the F1 scores obtained in each class and the macro and weighted average of these scores. 
    
    For CICIDS2017, when comparing \ac{EFC} with the other classifiers, it had the best results in eight of the thirteen classes. In addition, it ranked second best in the classes
	\textit{DDoS}, \textit{PortScan} and \textit{SSH-Patator}. In overall metrics, \ac{EFC} achieved $0.752 \pm 0.013$ of F1 macro average and $0.940 \pm 0.001$ of F1 weighted average, representing the best and third best results, respectively. In other words, EFC is the best classifier if we consider all classes with the same importance and the third best if we consider the majority classes as the most important. \begin{table}[!ht]
		\centering
		\input{tab4.tex}
	\end{table}
	
	Despite being significantly better than all other classifiers, \ac{EFC} performance in \textit{Bot}, \textit{Infiltration} and \textit{Web Attack} was far from good: $0.585$, $0.347$, and $0.574$ of F1 on average, respectively. These results can be explained by the \blue{fact} that the Precision pulls down the F1 scores and not the Recall (around $0.90$) for the three classes. Since these classes correspond to less than 0.002\%, 0.07\%, and 0.08\% of the test set, their Precision is expected to be heavily impacted by small portions of misclassifications of majority classes. In other words, we interpret these results as a consequence of class imbalance in the test set and not as a clear indication of \ac{EFC}'s ability to characterize these attacks. 
    
    According to the CICIOMT2024 results depicted in Table \ref{tab:compCICIoMT2024}, \ac{EFC} achieved the top score specifically for the \textit{\blue{DoS} SYN} attack. In contrast, \ac{DT} outperformed it across the remaining 17 attack types and benign classification. This broader success of \ac{DT} may be linked to the data reduction and balancing required for its execution (as well as for SVM and MLP). Nonetheless, \ac{EFC} proved a strong competitor, ranking second for \textit{Benign} classification and ten specific attacks: \textit{ARP Spoofing, DDOS ICMP, DDoS TCP, DoS ICMP, DoS Publish \blue{Flood}, DoS UDP, Malformed Data, OS Scan, Ping Sweep}, and \textit{Port Scan}. 
    
    In conclusion, EFC demonstrates significant capabilities as a multi-class classifier across the evaluated datasets. Its achievement of the highest F1 macro average on CICIDS2017 highlights its capacity for balanced performance, treating all classes with equal importance. Furthermore, EFC secured the top F1 score in numerous individual classes across both CICIDS2017 (eight classes) and CICIOMT2024 (\textit{\blue{DoS} SYN}). Its frequent ranking as the second-best method in many other categories underscores its competitiveness. While F1 scores for specific rare classes on CICIDS2017 were lower, the analysis attributes this primarily to test set imbalance impacting precision, whereas recall remained near 0.90, suggesting underlying detection resilience. The performance comparison on CICIOMT2024 also indicates EFC's operational characteristics may differ from methods reliant on specific data reduction or balancing steps. These collective findings position EFC as a method warranting further investigation and consideration for multi-class network traffic classification tasks. 
    
    \begin{figure}[!h]
		\centering
		\includegraphics[width=0.8\columnwidth]{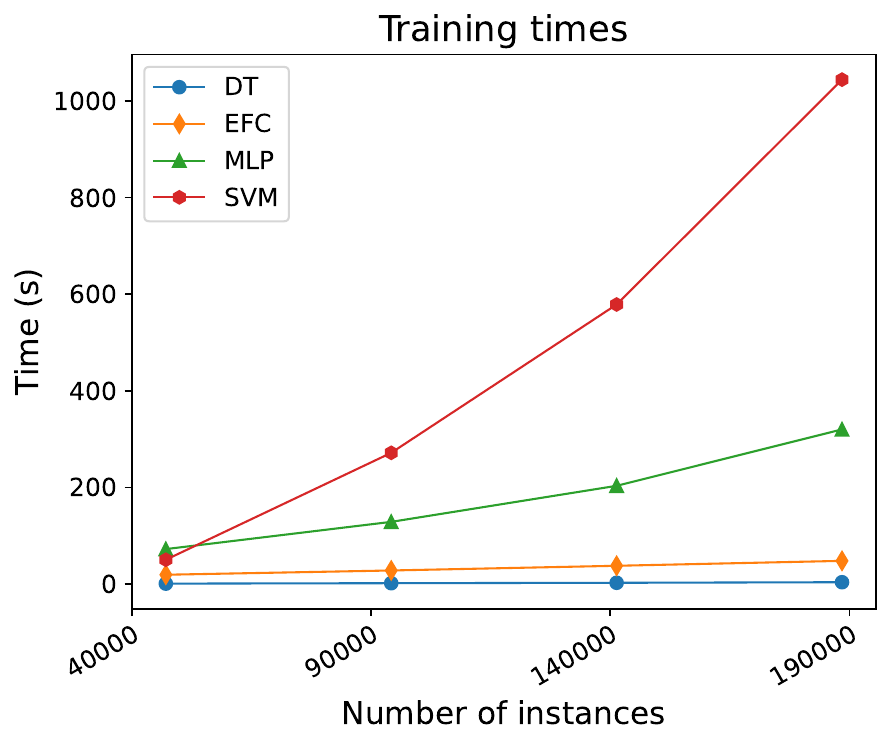}
		\caption{Training times of \ac{SVM}, \ac{MLP}, DT and \ac{EFC} algorithms for different training set sizes}
		\label{fig:times}
	\end{figure}
	
	We emphasize that the EFC has a low temporal complexity, significantly smaller than that of \ac{SVM} and \ac{MLP}, being comparable to the complexity of a \acl{DT}. Figure \ref{fig:times} shows the runtimes of these algorithms for different input sizes in \blue{an} Ubuntu 20.04.2 LTS OS with a 9th Generation Intel Core i5 processor, with 4 cores and hyperthreading. It is worth mentioning that to compare EFC with the literature solutions, OCN~\cite{Zhang2021} and ODIN~\cite{Liang2018}, we had to select an outdated version of operational system to cope with their requirements. Therefore, we selected an outdated version of \blue{the} operational system accordingly. \blue{The} empirical results show expected behavior, with the Decision Tree and \ac{EFC} growing linearly with the number of samples. However, empirical times should be interpreted with caution, as they depend heavily on the skills of the algorithm implementer. As Buczak \textit{et al.} \cite{Buczak2016} noted, \ac{SVM}, \ac{MLP}, and DT have well-maintained open-source implementations, which \blue{allow} them to achieve their best possible times. Meanwhile, \ac{EFC} is a method still under research whose implementation may not yet be the most efficient. Nevertheless, we consider that \ac{EFC} is a fast algorithm with a high applicability potential in real-time NIDS.
	\subsection{Open set comparative analysis}
	\label{subsec:fig}
	
	We perform open set comparative experiments considering \ac{EFC} and consolidated \ac{ML} algorithms, such as \ac{DT},  \ac{SVM}, and \ac{MLP}. Also, we present two experiments, considering CICIDS2017 \cite{Zhang2021} and CICIOMT2024 \cite{CICIoMT2024}.
	\subsubsection{CICIDS2017 Open Set Comparative Analysis}
	
	\begin{figure*}[ht]
		\centering
		\includegraphics[width=\textwidth]{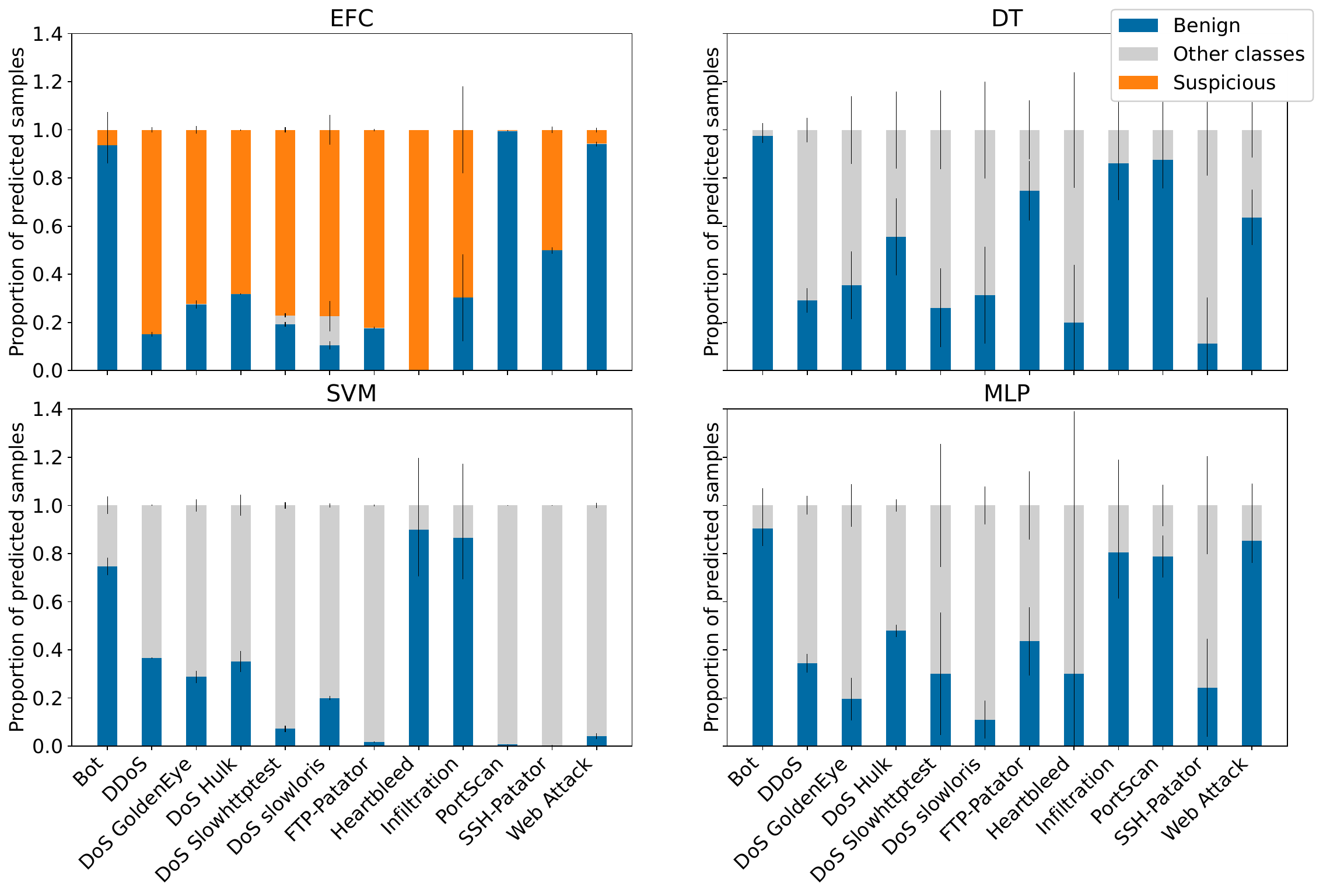}
		\caption{CICIDS2017 - Pictographic Classification of unknown attacks by \ac{EFC}, \ac{DT},  \ac{SVM}, and \ac{MLP} - Misclassifications represented in blue, desirable results are bars either orange for \ac{EFC} or gray for other classifiers.}
		\label{fig:unk1}
	\end{figure*}
	\begin{table}[!ht]\centering
		\centering
		\caption{CICIDS2017 Open-Set Experiment - Detailing the classification of unknown attacks by \ac{EFC}, \ac{DT},  \ac{SVM}, and \ac{MLP} - Misclassifications \blue{as the \textit{Benign} class are undesirable;} desirable results are classification as \textit{Suspicious} (for \ac{EFC}) or \textit{Other Attacks} (for other classifiers)\blue{(95\% CI)}.}\label{table:efc_os_legacy}
		\input{tab5}
	\end{table}
	The second evaluation in our work investigates the ability of classifiers to detect unknown attacks. To perform such an evaluation, we designed an open-set experiment where we turn a known attack into an unknown by removing it from the training set. Afterward, we assess the performance of our solution against the other algorithms considering a normal test set, which remains unchanged with samples of the original attack, evaluating the capability of both techniques to identify these samples as a threat to the network. Further details of this experiment can be found in Subsection \ref{sub:exp}. Next, we present its results.
    
    Figure \ref{fig:unk1} shows the result of each algorithm's classification of unknown samples. We also added Table \ref{table:efc_os_legacy} with the same results but numerically depicted. We believe that the pictographic chart is easier to understand the capabilities of EFC, compared to the other \ac{ML} algorithms. For example, the bar labeled \textit{Bot}, in Figure \ref{fig:unk1}, shows the classification of \textit{Bot} samples in the experiment where the \textit{Bot} class was omitted from training (making it an unknown attack). Note that the same reasoning can be applied to each \blue{row} of Table \ref{table:efc_os_legacy}. The colors in the bars represent the predicted classes of these samples, which can be \textit{Benign} if it was labeled as \textit{Benign}, or \textit{Other classes} if it was labeled as any other attack class. In addition, for the multi-class \ac{EFC}, we also have the \textit{Suspicious} class, which is the label provided by the classifier when samples do not fit into any known class, being classified as a possible threat to the network. The ideal results for \ac{SVM}, DT and \ac{MLP} would be full gray bars, which would mean that no unknown attack samples were classified as \textit{Benign}, even if they were misclassified. Meanwhile, for the \ac{EFC}, the ideal results would be full orange bars, which would mean that every attack was correctly recognized as an unknown attack. 
    
    From Figure \ref{fig:unk1}, we can see that \ac{EFC} classified the vast majority of unknown samples as \textit{suspicious} in almost all classes. The exceptions were the classes \textit{Bot}, \textit{PortScan}, and \textit{Web Attack}, which were almost totally misclassified, but were also challenging for most classifiers as they are application layer attacks and require deep packet inspection to be properly identified. 
    
    In this experiment, the \acl{DT} \blue{well-illustrates} the problem of unknown attack detection. Although it has good classification metrics as a multi-classifier and an excellent runtime speed, it does not properly identify new classes of attacks and mistakes them as benign samples. 
    
    From a network security point of view, in this experiment, the algorithm with the smallest FNR was SVM, as it is the least likely to classify attack samples as benign. However, the goal here is not just to flag attacks, like a binary classifier, but to distinguish and correctly classify. Not only SVM but also\blue{, apart from EFC,} the other algorithms misclassified the unknown attack, which would lead to an improper attack response. In the case of SVM, this clearly shows it is not a suitable algorithm. Not only does it not isolate unknown samples, but it also performs poorly as a multi-classifier. Additionally, it has a high computational cost, which hinders online detection. If a detector using those algorithms were to trigger some automated response, like in SOAR (Security Operations Automated Response), the effects of the mitigation might be ineffective or harmful, exacerbating those of the original attack itself. 
        
    In turn, EFC can identify unknown attacks as Suspicious for almost all classes, except PortScan, which is reconnaissance (not really an attack). The suspicious classification would indicate the need for further analysis, which would either confirm a new previously unclassified attack, some already known, or even a benign event. This certainly would mitigate the chance of an improper response. 
    
    For this reason, we consider \ac{EFC} a promising classifier to be used in \ac{NIDS}, as it combines high performance with temporal efficiency and the ability to detect unknown samples, potentially preventing improper incident response.
    
	\subsubsection{CICIOMT2024 Open Set Comparative Analysis}
	
	\begin{figure*}[ht]
		\centering
		\includegraphics[width=\textwidth]{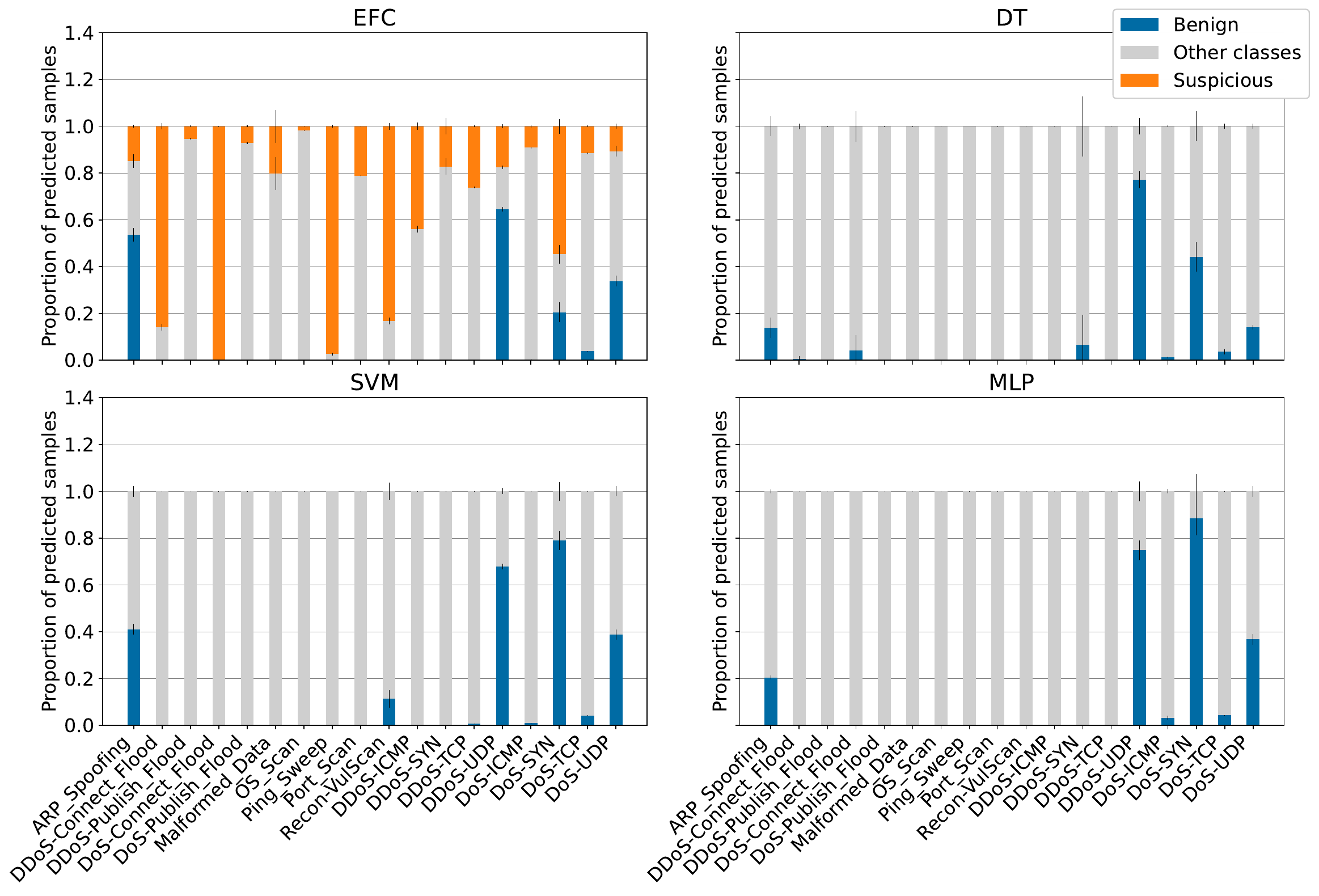}
		\caption{CICIOMT2024 - Pictographic classification of unknown attacks by \ac{EFC}, \ac{DT},  \ac{SVM}, and \ac{MLP} - Misclassifications represented in blue, desirable results are bars either orange for \ac{EFC} or gray for other classifiers.}
		\label{fig:ciciomt_unk1}
	\end{figure*}

	Following the same rationale as the previous experiment, we performed an open-set evaluation using the CICIOMT2024 dataset from Dadkhah \etal \cite{CICIoMT2024}. In this setup, we iteratively excluded one of the dataset's 18 classes during model training, treating this excluded class as an 'unknown' attack. During the test phase, samples from this 'unknown' attack class were used for evaluation. This entire procedure was repeated for each of the 18 classes acting as the unknown attack, and was applied to all tested \ac{ML} algorithms, including EFC. Figure \ref{fig:ciciomt_unk1} depicts the results for each algorithm across these scenarios, using stacked bars to provide a comprehensive assessment for each tested unknown attack. In addition, Table \ref{tab:ciciomt_unk1} presents the raw values obtained to create Figure \ref{fig:ciciomt_unk1}. \begin{table}[!ht]\centering
		\centering
		\caption{CICIOMT2024 Open-Set Experiment - Detailing the classification of unknown attacks by \ac{EFC}, \ac{DT},  \ac{SVM}, and \ac{MLP} - Misclassifications \blue{as the \textit{Benign} class are undesirable;} desirable results are classification as \textit{Suspicious} (for \ac{EFC}) or \textit{Other Attacks} (for other classifiers). \blue{(95\% CI)}}\label{tab:ciciomt_unk1}
		\input{tab6}
	\end{table}
	
	In this open-set evaluation, \blue{the Decision Tree (DT) had the lowest rate of misclassifying unknown attacks as 'Benign' (False Negatives)}. DT consistently demonstrated the lowest rate of misclassifying unknown attacks as 'Benign' (False Negatives) compared to EFC, SVM, and MLP. This indicates it had the highest True Positive Rate for recognizing threats, correctly identifying many unknown attack types like DDoS Publish Flood, Malformed Data, OS Scan, and Ping Sweep as malicious ('Other Attacks') nearly 100\% of the time. Its overall lower sum of 'Benign' classification rates across all tested unknown classes reinforces its superior performance in distinguishing malicious traffic in this scenario. However, it is important to note that although \ac{DT} had the best performance to separate benign from malicious samples, the classification was\blue{,} in fact, a misclassification given that the unknown attack was attributed to an incorrect class of attack, whereas EFC classified \blue{it} as suspicious. 
        
    EFC, in turn, demonstrates a valuable capability in identifying certain types of unknown attacks specifically as 'Suspicious'. It showed good performance in flagging novel threats like DoS Connect Flood (classifying it as suspicious 99.9\% of the time), Ping Sweep (97.4\%), DDoS Connect Flood (85.9\%), and Recon VulScan (83.2\%). While its performance varies across different attack types, EFC's ability to successfully isolate and label these specific unknown activities as suspicious highlights its potential as a tool for detecting novel threats within network traffic. 
    
    Conversely, the Support Vector Machine (SVM) technique made the most mistakes, exhibiting the highest overall tendency to misclassify unknown attacks as 'Benign' (False Negatives). This means it was the least successful at detecting novel threats. SVM showed particularly high error rates for certain attack types, such as classifying \blue{DoS} SYN as benign 79.1\% of the time and DDoS UDP as benign 67.9\% of the time. Its highest overall sum of average 'Benign' rates across all unknown classes highlights its relative weakness in this open-set test compared to the other evaluated methods.
	
    \subsection{Open set literature comparison}
    
	 In this section, we evaluate EFC against the literature solutions considering open set experiments. First, we compared our solution with the OCN from Zhang \textit{et al.}~\cite{Zhang2021} and ODIN \cite{Liang2018}. Afterwards, given the superiority of OCN against ODIN, we evaluate EFC and OCN in an open set experiment considering the CICIOMT2024 from Dadkhah \etal \cite{CICIoMT2024} dataset.
	\subsubsection{CICIDS2017  Open Set Literature Comparison}
	\begin{table}[ht]
		\centering
		\input{tab7}
	\end{table}
	Our third experiment was a reproduction of Zhang \textit{et al.}\blue{'s} assessment of open set classifiers. The authors compared their proposal  OCN \cite{Zhang2021} with a Baseline and ODIN \cite{Liang2018} open set techniques. We reused their performance results to compare with the Multi-class EFC, considering the exact same dataset \blue{and} following their instructions for reproducibility. It is worth mentioning that we had to reuse performance results from the work \blue{by Zhang et al.}~\cite{Zhang2021}, since the Baseline, OCN \cite{Zhang2021}\blue{,} and ODIN \cite{Liang2018}, presented performance below \blue{that} reported. The evaluation design is similar to the experiment in section \ref{subsec:closed}, where one type of attack at a time was dropped from the training set. Table \ref{tab:unk2} shows the results of this comparison for each algorithm in terms of the AUROC and AUPRC metrics. 
    
    The EFC recognized virtually all unknown samples of \textit{Heartbleed}, \textit{Infiltration}, and \textit{Botnet}, with both AUROC and AUPRC above 0.99. It outperformed all other algorithms in these last two classes and matched the OCN in the first. In \textit{DoS slowloris}, \textit{Web Attack}, and \textit{DoS \blue{Slowhttptest}}, OCN obtained the best AUROC values, showing that they can identify both known and unknown samples with greater recall than the EFC. However, in the same classes, EFC obtained the best AUPRC, suggesting that it is the least likely to classify unknown samples as known. These results \blue{show} that EFC, although not always superior, is at least similar to the state of the art in detecting unknown attacks. 
    
    It is important to note that in this experiment, we use the features extracted by the authors of \cite{Zhang2021}, which are ideal for their method but may negatively affect the performance of EFC. Evidence of this is that when using the original dataset, around 80\% of \textit{DoS slowloris} and \textit{DoS \blue{Slowhttptest}} samples were correctly recognized by EFC  (Table \ref{tab:compCICIDS2017}), while in this experiment, they were not well recognized.

	\subsubsection{CICIOMT2024 Open Set Literature Comparison}
	
	\begin{table}[!ht]
		\centering
\input{tab8}

	\end{table}
	
	The last experiment performed compares EFC against OCN considering the dataset from Dadkhah \etal \cite{CICIoMT2024}. The dataset CICIOMT2024 was randomly sampled and reduced to present at most 5000 samples per class, as presented in Table \ref{table:compCICIOMT2024}. First, the attack had its samples removed from the training phase for each class. Afterwards, we added it back for the test, keeping the original ratio of 20\% of the dataset reserved for test and 80\% for training \cite{Zhang2021}. We collected the metrics AUROC and AUPRC for both solutions and \blue{present them} in Table \ref{tab:unk2-2024}. 
    
    As can be noted, EFC outperformed in classification regarding AUROC for 9 out of the 18 attacks considered, namely \textit{ARP Spoofing, DDoS ICMP, DDoS SYN, DDoS UDP, DoS ICMP, DoS SYN, DoS TCP, DoS UDP,} and \textit{VulScan}. Whereas, OCN \cite{Zhang2021} outperformed EFC in \textit{DDoS Connect Flood, DDoS Publish Flood, DDoS TCP, DoS Connect Flood, DoS Publish Flood, Malformed Data, OS Scan, Ping Sweep,} and \textit{Port Scan}. Different from the results for CICIDS2017 \blue{from} Zhang \etal \cite{Zhang2021}, EFC performed better than OCN regarding AUPRC only for the attacks \textit{DDoS UDP} and \textit{DoS TCP}. These results show how sensitive EFC is regarding different dataset preparations. Therefore, EFC's performance may be degraded by the dataset preprocessing procedure.
	\section{Conclusion}
	\label{sec:conclusion}
	
	In this work, we proposed a new open-set multi-class classifier for NIDS: the multi-class EFC. Our method is an extension of the single-class EFC, first introduced in \cite{pontes2019new}, that performs classification with respect to a benign class, several attack classes, and a suspicious class (intended for unknown attacks). We evaluated our proposal using the CICIDS2017 and CICIOMT2024 datasets, which include a range of modern network attacks. Other studies also used these datasets, allowing us to compare our performance against existing state-of-the-art results directly. 
    
    In our first experiment, we performed cross-validation and compared EFC with well-established machine learning classifiers such as SVM, DT, and MLP. From this evaluation, we conclude that the EFC is better than the others in eight of the thirteen classes of attacks, reaching the best value of the F1 macro average. In addition, we highlight the low temporal complexity of our proposal and empirically show that it is a fast algorithm. 
    
    In the second part of our evaluation, we investigated EFC's mechanism to identify unknown attacks by comparing it with the classical algorithms from the previous experiment and other open-set algorithms from the literature in open-set experiments. These results showed that the mechanism is effective and that EFC has a detection performance for unknown samples that is similar or comparable to the state-of-the-art. It must be noted that the setup used for EFC was not optimized, following \blue{state-of-the-art} dataset preparation procedures, so \blue{EFC's} performance may be further enhanced. 
    
    In the future, we intend to investigate a dynamic threshold to replace the static 95th percentile used in the current implementation. We are developing a real-time EFC version integrated with Software-Defined Networks. This implementation will allow us to evaluate the EFC's performance at line-rate and compare it against other real-time solutions reported in the literature.
	\section*{Acknowledgment}
The authors wish to acknowledge the invaluable contributions of Mr. Álvaro Veloso Cavalcanti Luz towards the completion of this work. Prof. Marcelo A. Marotta was financially supported by the project PROFISSA - Programmable Future Internet for Secure Software Architectures under grant 2020/05152-7, São Paulo Research Foundation (FAPESP). Prof. Luiz DaSilva is partially supported by the Commonwealth Cyber Initiative.
	\bibliographystyle{elsarticle-num} 
	
	\bibliography{bibliography.bib}
	
	\section*{Biographies}
	
	\textbf{Manuela M. C. de Souza }is a graduate student at the Federal University of Pernambuco (UFPE), focusing on embedded software for the avionics industry. She earned her Bachelor's in Computer Science from the University of Brasília (UnB), engaging in research on computer security and machine learning. Her primary research interests include Network Security, Machine Learning, and Embedded Software.
    
	\textbf{Camila F. T. Pontes }works as a researcher at the Barcelona Supercomputing Center (BSC) in Spain. She holds undergraduate degrees in Biology (2014) and Computer Science (2020) from the University of Brasilia. From the same university, she obtained her M.Sc. (2016) and Ph.D. (2021) in Computational Biology. Her research focuses on Computational and Theoretical Biology and Network Security.
	
	\textbf{João J. C. Gondim} earned his M.Sc. in Computing Science from Imperial College, University of London (1987), and a Ph.D. in Electrical Engineering from the University of Brasilia (UnB, 2017). He is an associate professor and tenured faculty member in the Department of Electrical Engineering (ENE) at UnB. His research interests encompass network, information, and cybernetic security.
	
	\textbf{Luís Paulo Faina Garcia} holds a degree in Computer Engineering (2010) and a Ph.D. in Computer Science (2016) from the University of São Paulo. His doctoral thesis was recognized as one of the best by the Brazilian Computer Society in 2017 and won the CAPES award for the best Computer Science thesis nationally. Currently an Assistant Professor at the University of Brasília's Department of Computer Science, his expertise covers noise detection, meta-learning, and data streams.

    \textbf{Luiz A. DaSilva} serves as the Executive Director of the Commonwealth Cyber Initiative and holds the Bradley Professor of Cybersecurity position at Virginia Tech. Formerly, he was at Trinity College Dublin, directing CONNECT, the Science Foundation Ireland Research Centre for Future Networks. He is distinguished as an IEEE Fellow and an IEEE Communications Society Distinguished Lecturer.

    \textbf{Eduardo Ferreira Marques Cavalcante} is pursuing an undergraduate degree in Computer Science at the University of Brasilia. He aims to specialize in Computer Vision and Artificial Intelligence. Additionally, he possesses experience in software engineering and web development.
	
    \textbf{Marcelo Antonio Marotta} is an assistant professor and tenured faculty member at the University of Brasilia. He obtained his Ph.D. in Computer Science in 2019 from the Institute of Informatics (INF) at the Federal University of Rio Grande do Sul (UFRGS), Brazil. His research areas include Heterogeneous Cloud Radio Access Networks, Internet of Things, Software Defined Radio, Cognitive Radio Networks, and Network Security.
\end{document}

%% file: tab1.tex
\renewcommand{\arraystretch}{1.5}
\resizebox{0.75\columnwidth}{!}{%
\begin{tabular}{l r r r r}
    \hline
    Flow class & & \multicolumn{3}{c}{Number of instances}\\
    \cline{3-5}
    & & Original CICIDS2017 & & Zhang \textit{et al.} CICIDS2017\\
    \hline
    \textit{BENIGN} & & 2,272,688 & & 62,639\\
    \textit{DDoS} & & 128,027& & 261,226\\
    \textit{PortScan} & & 158,930& & 319,636\\
    \textit{FTP-Patator} & & 7,938& & 19,941\\
    \textit{DoS Hulk} & & 23,0124& & 474,656\\
    \textit{DoS GoldenEye} & & 10,293& & 20,543\\
    \textit{DoS slowloris} & & 5,796& & 10,537\\
    \textit{Bot} & & 1,966& & 2,075\\
    \textit{Infiltration} & & 36& & 5,330\\
    \textit{Web Attack} & & 2,180& & 10,537\\
    \textit{SSH-Patator} & & 5,897& & 27,545\\
    \textit{DoS Slowhttptest} & & 5,499& & 6,786\\
    \textit{Heartbleed} & & 11& & 9,859\\
    \hline
\end{tabular}
}

%% file: tab2.tex
\resizebox{0.75\columnwidth}{!}{%
\begin{tabular}{l r r r r}
\hline
Flow class & & \multicolumn{3}{c}{Number of instances}\\
    \cline{3-5}
    & & Original  & & After Random Sampling \\
    
\hline
\textit{Benign} & & 192732 & & 5000 \\
\textit{ARP Spoofing} & & 16046 & & 5000\\
\textit{DDoS Connect Flood} & & 173036  & & 5000 \\
\textit{DDoS ICMP} & & 1537476 & & 5000 \\
\textit{DDoS Publish Flood} & & 27623  & & 5000\\
\textit{DDoS SYN} & & 801962  & & 5000\\
\textit{DDoS TCP} & & 804465  & & 5000\\
\textit{DDoS UDP} & & 1635956  & & 5000\\
\textit{DoS Connect Flood} & & 12773  & & 5000\\
\textit{DoS ICMP} & & 416292 & & 5000 \\
\textit{DoS Publish Flood} & & 44376  & & 5000\\
\textit{DoS SYN} & & 441903 & & 5000 \\
\textit{DoS TCP} & & 380384  & & 5000\\
\textit{DoS UDP} & & 566950  & & 5000\\
\textit{Malformed Data} & & 5130  & & 5000\\
\textit{OS Scan} & & 16832  & & 5000\\
\textit{Ping Sweep} & & 740  & & 740\\
\textit{Port Scan} & & 83981  & & 5000\\
\textit{VulScan} & & 2173  & & 2130\\
\hline
\end{tabular}
}

%% file: tab3.tex
\caption{CICIDS2017 - Average classification performance and standard error (95\%~CI)}
\label{tab:compCICIDS2017}
\renewcommand{\arraystretch}{1.5}
\resizebox{0.9\columnwidth}{!}{%
\begin{tabular}{l c c c c c c c c}
\hline
Class & & EFC & & DT & & SVM & & MLP\\
\hline
%quinto, RF em primeiro
\textit{BENIGN} & & 0.949 $\pm$ 0.001 & & \textbf{0.994 $\pm$ 0.000}& & 0.902 $\pm$ 0.003& & 0.978 $\pm$ 0.003\\ 
\textit{Bot} & &\textbf{0.585 $\pm$ 0.019} & & 0.544 $\pm$ 0.065& & 0.031 $\pm$ 0.003& & 0.501 $\pm$ 0.046\\ 
\textit{DDoS} & & 0.966 $\pm$ 0.002 & &  \textbf{0.992 $\pm$ 0.002}& & 0.749 $\pm$ 0.002& & 0.964 $\pm$ 0.012\\ 
\textit{DoS GoldenEye} & & \textbf{0.967 $\pm$ 0.002} & & 0.895 $\pm$ 0.030& & 0.437 $\pm$ 0.024& & 0.874 $\pm$ 0.010\\ 
\textit{DoS Hulk} & & 0.823 $\pm$ 0.005 & & \textbf{0.991 $\pm$ 0.001}& & 0.897 $\pm$ 0.009& & 0.952 $\pm$ 0.011\\ 
\textit{DoS Slowhttptest} & & \textbf{0.917 $\pm$ 0.008} & & 0.832 $\pm$ 0.029& & 0.752 $\pm$ 0.012& & 0.799 $\pm$ 0.029\\ 
\textit{DoS slowloris} & & \textbf{0.963 $\pm$ 0.003} & &  0.786 $\pm$ 0.037& & 0.505 $\pm$ 0.053& & 0.818 $\pm$ 0.040\\ 
\textit{FTP-Patator} & & \textbf{0.974 $\pm$ 0.004} & & 0.932 $\pm$ 0.023& & 0.454 $\pm$ 0.041& & 0.843 $\pm$ 0.033\\ 
\textit{Heartbleed} & & \textbf{0.800 $\pm$ 0.160} & & 0.055 $\pm$ 0.029& & 0.302 $\pm$ 0.345& & 0.336 $\pm$ 0.191 \\ 
\textit{Infiltration} & & \textbf{0.347 $\pm$ 0.171} & & 0.018 $\pm$ 0.007& & 0.069 $\pm$ 0.019& & 0.067 $\pm$ 0.038\\ 
\textit{PortScan} & &  0.969 $\pm$ 0.002 & & \textbf{0.987 $\pm$ 0.002}& & 0.863 $\pm$ 0.028& & 0.901 $\pm$ 0.002\\ 
\textit{SSH-Patator} & & 0.701 $\pm$ 0.041 & & \textbf{0.961 $\pm$ 0.039}& & 0.191 $\pm$ 0.004& & 0.664 $\pm$ 0.076  \\ 
\textit{Web Attack} & & \textbf{0.574 $\pm$ 0.097} & & 0.520 $\pm$ 0.048& & 0.216 $\pm$ 0.004& & 0.307 $\pm$ 0.015  \\ 
\hline
\textbf{Macro average} & & \textbf{0.752 $\pm$ 0.013} & &  0.731 $\pm$ 0.006& & 0.490 $\pm$ 0.026& & 0.693 $\pm$ 0.017 \\ 
\textbf{Weighted average} & & 0.940 $\pm$ 0.001 & & \textbf{0.991 $\pm$ 0.001}& & 0.886 $\pm$ 0.003& & 0.968 $\pm$ 0.004  \\ 
\hline
\end{tabular}
}

%% file: tab4.tex
\caption{CICIOMT2024 - Average classification performance and standard error (95\%~CI)}
\label{tab:compCICIoMT2024}
\renewcommand{\arraystretch}{1.5}
\resizebox{0.9\columnwidth}{!}{%
\begin{tabular}{l c c c c c c c c}
\hline
Class & & EFC & & DT & & SVM & & MLP\\
\hline
%quinto, RF em primeiro
\textit{Benign}  & & 0.477 $\pm$ 0.005 & & \textbf{0.732 $\pm$ 0.011 }& & 0.363 $\pm$ 0.010 & & 0.438 $\pm$ 0.014\\ 
\textit{ARP Spoofing}  & & 0.917 $\pm$ 0.001 & & \textbf{0.954 $\pm$ 0.002} & & 0.888 $\pm$ 0.005 & & 0.909 $\pm$ 0.004\\
\textit{DDoS Connect Flood}  & & 0.971 $\pm$ 0.002 & & \textbf{1.000 $\pm$ 0.000} & & 0.981 $\pm$ 0.001 & & 0.991 $\pm$ 0.002\\
\textit{DDoS ICMP}  & & 0.935 $\pm$ 0.010 & & \textbf{0.997 $\pm$ 0.001} & & 0.624 $\pm$ 0.002 & & 0.726 $\pm$ 0.023\\
\textit{DDoS Publish Flood}  & & 0.943 $\pm$ 0.002 & & \textbf{0.999 $\pm$ 0.000} & & 0.761 $\pm$ 0.209 & & 0.506 $\pm$ 0.181\\
\textit{DDoS SYN}  & & 0.942 $\pm$ 0.011 & & \textbf{0.998 $\pm$ 0.000} & & 0.276 $\pm$ 0.170 & & 0.285 $\pm$ 0.197\\
\textit{DDoS TCP}  & & 0.849 $\pm$ 0.023 & & \textbf{0.996 $\pm$ 0.001} & & 0.244 $\pm$ 0.002 & & 0.549 $\pm$ 0.007\\
\textit{DDoS UDP}  & & 0.350 $\pm$ 0.006 & & \textbf{0.681 $\pm$ 0.012} & & 0.352 $\pm$ 0.005 & & 0.426 $\pm$ 0.024\\
\textit{DoS Connect Flood}  & & 0.934 $\pm$ 0.003 & & 0.988 $\pm$ 0.007 & & 0.977 $\pm$ 0.004 & & \textbf{0.991 $\pm$ 0.003}\\
\textit{DoS ICMP}  & & 0.430 $\pm$ 0.023 & & \textbf{0.682 $\pm$ 0.008} & & 0.251 $\pm$ 0.017 & & 0.297 $\pm$ 0.007\\
\textit{DoS Publish Flood}  & & 0.949 $\pm$ 0.004 & & \textbf{0.998 $\pm$ 0.000} & & 0.820 $\pm$ 0.001 & & 0.843 $\pm$ 0.055\\
\textit{DoS SYN}  & & \textbf{0.633 $\pm$ 0.017} & & 0.542 $\pm$ 0.022 & & 0.351 $\pm$ 0.017 & & 0.474 $\pm$ 0.039\\
\textit{DoS TCP}  & & 0.673 $\pm$ 0.039 & & \textbf{0.930 $\pm$ 0.002} & & 0.814 $\pm$ 0.028 & & 0.699 $\pm$ 0.154\\
\textit{DoS UDP}  & & 0.372 $\pm$ 0.009 & & \textbf{0.650 $\pm$ 0.016} & & 0.314 $\pm$ 0.014 & & 0.357 $\pm$ 0.021\\
\textit{Malformed Data}  & & 0.941 $\pm$ 0.009 & & \textbf{0.998 $\pm$ 0.000} & & 0.531 $\pm$ 0.306 & & 0.668 $\pm$ 0.200\\
\textit{OS Scan}  & & 0.936 $\pm$ 0.012 & & \textbf{0.999 $\pm$ 0.000} & & 0.835 $\pm$ 0.001 & & 0.781 $\pm$ 0.028\\
\textit{Ping Sweep}  & & 0.963 $\pm$ 0.004 & & \textbf{0.988 $\pm$ 0.004} & & 0.796 $\pm$ 0.004 & & 0.884 $\pm$ 0.024\\
\textit{Port Scan}  & & 0.878 $\pm$ 0.007 & & \textbf{0.997 $\pm$ 0.001} & & 0.269 $\pm$ 0.057 & & 0.376 $\pm$ 0.046\\
\textit{VulScan}  & & 0.946 $\pm$ 0.002 & & \textbf{0.995 $\pm$ 0.001} & & 0.982 $\pm$ 0.003 & & 0.994 $\pm$ 0.001\\
\hline
\textbf{Macro average}  & & 0.752 $\pm$ 0.005 & & \textbf{0.901 $\pm$ 0.002} & & 0.601 $\pm$ 0.016 & & 0.642 $\pm$ 0.006\\ 
\textbf{Weighted average}  & & 0.924 $\pm$ 0.006 & & \textbf{0.995 $\pm$ 0.000} & & 0.664 $\pm$ 0.063 & & 0.652 $\pm$ 0.020\\ 
\hline
\end{tabular}}

%% file: tab5.tex
\begin{sideways}
\resizebox{1.18\columnwidth}{!}{%
\begin{tabular}{lcccccccccccccccccc}
\hline
 & & \multicolumn{5}{c}{EFC} & & \multicolumn{3}{c}{DT} & & \multicolumn{3}{c}{SVM} & & \multicolumn{3}{c}{MLP} \\
\cline{3-7}
\cline{9-11}
\cline{13-15}
\cline{17-19}
Flow class    & & Benign & &Other Attacks & &Suspicious & &Benign & &Other Attacks & &Benign & &Other Attacks & &Benign & &Other Attacks  \\
\hline    
Bot & & 0.935  $\pm$ 0.075 & & 0.000  $\pm$ 0.001 & &0.065  $\pm$ 0.076 & & 0.976  $\pm$ 0.031 & & 0.055  $\pm$ 0.028 & &0.749 $\pm$ 0.037 & & 0.251  $\pm$ 0.038 & &0.906  $\pm$ 0.075 && 0.094  $\pm$ 0.076 \\
DdoS  & & 0.154  $\pm$ 0.012 & & 0.000  $\pm$ 0.001 & & 0.846  $\pm$ 0.010 & & 0.290  $\pm$ 0.053 & & 0.763  $\pm$ 0.050 & & 0.367  $\pm$ 0.003 & & 0.633  $\pm$ 0.004 & & 0.346  $\pm$ 0.040 & & 0.654  $\pm$ 0.041 \\
DoS Golden Eye& & 0.278 $\pm$ 0.019 & & 0.003 $\pm$ 0.001 & & 0.719 $\pm$ 0.016 & & 0.355 $\pm$ 0.144 & & 0.789 $\pm$ 0.140 & & 0.287 $\pm$ 0.023 & & 0.713 $\pm$ 0.027 & & 0.198 $\pm$ 0.091 & & 0.802 $\pm$ 0.090 \\
DoS Hulk         & & 0.319 $\pm$ 0.002 & & 0.000 $\pm$ 0.001 & & 0.681 $\pm$ 0.002 & & 0.555 $\pm$ 0.163 & & 0.608 $\pm$ 0.160 & & 0.351 $\pm$ 0.042 & & 0.649 $\pm$ 0.046 & & 0.481 $\pm$ 0.029 & & 0.519 $\pm$ 0.031 \\
DoS Slowhttptest & & 0.194 $\pm$ 0.010 & & 0.035 $\pm$ 0.008 & & 0.771 $\pm$ 0.012 & & 0.260 $\pm$ 0.164 & & 0.904 $\pm$ 0.162 & & 0.074 $\pm$ 0.016 & & 0.926 $\pm$ 0.016 & & 0.301 $\pm$ 0.255 & & 0.699 $\pm$ 0.257 \\
DoS Slowloris    & & 0.108 $\pm$ 0.018 & & 0.118 $\pm$ 0.064 & & 0.774 $\pm$ 0.061 & & 0.313 $\pm$ 0.202 & & 0.889 $\pm$ 0.201 & & 0.201 $\pm$ 0.009 & & 0.799 $\pm$ 0.012 & & 0.112 $\pm$ 0.080 & & 0.888 $\pm$ 0.082 \\
FTP-Patator      & & 0.179 $\pm$ 0.005 & & 0.000 $\pm$ 0.001 & & 0.821 $\pm$ 0.005 & & 0.748 $\pm$ 0.126 & & 0.377 $\pm$ 0.122 & & 0.017 $\pm$ 0.004 & & 0.983 $\pm$ 0.007 & & 0.437 $\pm$ 0.143 & & 0.563 $\pm$ 0.143 \\
Heartbleed       & & 0.004 $\pm$ 0.001 & & 0.000 $\pm$ 0.001 & & 0.996 $\pm$ 0.000 & & 0.200 $\pm$ 0.241 & & 0.559 $\pm$ 0.238 & & 0.900 $\pm$ 0.199 & & 0.100 $\pm$ 0.198 & & 0.302 $\pm$ 0.393 & & 0.698 $\pm$ 0.394 \\
Infiltration     & & 0.303 $\pm$ 0.179 & & 0.000 $\pm$ 0.001 & & 0.697 $\pm$ 0.181 & & 0.862 $\pm$ 0.156 & & 0.294 $\pm$ 0.152 & & 0.865 $\pm$ 0.175 & & 0.135 $\pm$ 0.174 & & 0.805 $\pm$ 0.192 & & 0.195 $\pm$ 0.192 \\
PortScan         & & 0.995 $\pm$ 0.001 & & 0.000 $\pm$ 0.001 & & 0.005 $\pm$ 0.000 & & 0.873 $\pm$ 0.116 & & 0.242 $\pm$ 0.114 & & 0.008 $\pm$ 0.002 & & 0.992 $\pm$ 0.006 & & 0.790 $\pm$ 0.088 & & 0.210 $\pm$ 0.088 \\
SSH-Patator      & & 0.499 $\pm$ 0.014 & & 0.001 $\pm$ 0.001 & & 0.500 $\pm$ 0.014 & & 0.113 $\pm$ 0.190 & & 0.698 $\pm$ 0.187 & & 0.004 $\pm$ 0.002 & & 0.996 $\pm$ 0.005 & & 0.243 $\pm$ 0.203 & & 0.757 $\pm$ 0.207 \\
Web Attack       & & 0.943 $\pm$ 0.013 & & 0.001 $\pm$ 0.010 & & 0.056 $\pm$ 0.010 & & 0.635 $\pm$ 0.115 & & 0.480 $\pm$ 0.114 & & 0.042 $\pm$ 0.013 & & 0.958 $\pm$ 0.016 & & 0.854 $\pm$ 0.092 & & 0.146 $\pm$ 0.091 \\
\hline
\end{tabular}}
\end{sideways}

%% file: tab6.tex
\begin{sideways}
\resizebox{1.18\columnwidth}{!}{%
\begin{tabular}{lcccccccccccccccccc}
\hline
 & & \multicolumn{5}{c}{EFC} & & \multicolumn{3}{c}{DT} & & \multicolumn{3}{c}{SVM} & & \multicolumn{3}{c}{MLP} \\
\cline{3-7}
\cline{9-11}
\cline{13-15}
\cline{17-19}
Unknown class    & & Benign & &Other Attacks & &Suspicious & &Benign & &Other Attacks & &Benign & &Other Attacks & &Benign & &Other Attacks  \\
\hline    
ARP Spoofing & & 0.537 $\pm$ 0.029 & & 0.315 $\pm$ 0.029 & & 0.148 $\pm$ 0.006 & & 0.139 $\pm$ 0.043 & & 0.861 $\pm$ 0.043 & & 0.410 $\pm$ 0.023 & & 0.590 $\pm$ 0.023 & & 0.204 $\pm$ 0.008 & & 0.796 $\pm$ 0.008 \\
DDoS Connect Flood & & 0.000 $\pm$ 0.000 & & 0.141 $\pm$ 0.014 & & 0.859 $\pm$ 0.014 & & 0.006 $\pm$ 0.012 & & 0.994 $\pm$ 0.012 & & 0.000 $\pm$ 0.000 & & 1.000 $\pm$ 0.000 & & 0.000 $\pm$ 0.000 & & 1.000 $\pm$ 0.000 \\
DDoS Publish Flood & & 0.000 $\pm$ 0.000 & & 0.947 $\pm$ 0.003 & & 0.053 $\pm$ 0.003 & & 0.000 $\pm$ 0.000 & & 1.000 $\pm$ 0.000 & & 0.000 $\pm$ 0.000 & & 1.000 $\pm$ 0.000 & & 0.000 $\pm$ 0.000 & & 1.000 $\pm$ 0.000 \\
DoS Connect Flood & & 0.000 $\pm$ 0.000 & & 0.001 $\pm$ 0.000 & & 0.999 $\pm$ 0.000 & & 0.042 $\pm$ 0.066 & & 0.958 $\pm$ 0.066 & & 0.001 $\pm$ 0.000 & & 0.999 $\pm$ 0.000 & & 0.000 $\pm$ 0.000 & & 1.000 $\pm$ 0.000 \\
DoS Publish Flood & & 0.000 $\pm$ 0.000 & & 0.928 $\pm$ 0.004 & & 0.072 $\pm$ 0.004 & & 0.000 $\pm$ 0.000 & & 1.000 $\pm$ 0.000 & & 0.000 $\pm$ 0.000 & & 1.000 $\pm$ 0.000 & & 0.000 $\pm$ 0.000 & & 1.000 $\pm$ 0.000 \\
Malformed Data & & 0.000 $\pm$ 0.000 & & 0.797 $\pm$ 0.070 & & 0.203 $\pm$ 0.070 & & 0.000 $\pm$ 0.000 & & 1.000 $\pm$ 0.000 & & 0.000 $\pm$ 0.000 & & 1.000 $\pm$ 0.000 & & 0.000 $\pm$ 0.000 & & 1.000 $\pm$ 0.000 \\
OS Scan & & 0.000 $\pm$ 0.000 & & 0.982 $\pm$ 0.001 & & 0.018 $\pm$ 0.001 & & 0.000 $\pm$ 0.000 & & 1.000 $\pm$ 0.000 & & 0.000 $\pm$ 0.000 & & 1.000 $\pm$ 0.000 & & 0.000 $\pm$ 0.000 & & 1.000 $\pm$ 0.000 \\
Ping Sweep & & 0.000 $\pm$ 0.000 & & 0.026 $\pm$ 0.006 & & 0.974 $\pm$ 0.006 & & 0.000 $\pm$ 0.000 & & 1.000 $\pm$ 0.000 & & 0.000 $\pm$ 0.000 & & 1.000 $\pm$ 0.000 & & 0.000 $\pm$ 0.000 & & 1.000 $\pm$ 0.000 \\
Port Scan & & 0.000 $\pm$ 0.000 & & 0.789 $\pm$ 0.002 & & 0.211 $\pm$ 0.002 & & 0.000 $\pm$ 0.000 & & 1.000 $\pm$ 0.000 & & 0.001 $\pm$ 0.000 & & 0.999 $\pm$ 0.000 & & 0.000 $\pm$ 0.000 & & 1.000 $\pm$ 0.000 \\
Recon VulScan & & 0.002 $\pm$ 0.001 & & 0.166 $\pm$ 0.015 & & 0.832 $\pm$ 0.015 & & 0.000 $\pm$ 0.000 & & 1.000 $\pm$ 0.000 & & 0.114 $\pm$ 0.038 & & 0.886 $\pm$ 0.038 & & 0.000 $\pm$ 0.000 & & 1.000 $\pm$ 0.000 \\
DDoS ICMP & & 0.000 $\pm$ 0.000 & & 0.561 $\pm$ 0.015 & & 0.439 $\pm$ 0.015 & & 0.001 $\pm$ 0.001 & & 0.999 $\pm$ 0.001 & & 0.000 $\pm$ 0.000 & & 1.000 $\pm$ 0.000 & & 0.000 $\pm$ 0.000 & & 1.000 $\pm$ 0.000 \\
DDoS SYN & & 0.000 $\pm$ 0.000 & & 0.828 $\pm$ 0.034 & & 0.172 $\pm$ 0.034 & & 0.066 $\pm$ 0.129 & & 0.934 $\pm$ 0.129 & & 0.000 $\pm$ 0.000 & & 1.000 $\pm$ 0.000 & & 0.000 $\pm$ 0.000 & & 1.000 $\pm$ 0.000 \\
DDoS TCP & & 0.000 $\pm$ 0.000 & & 0.738 $\pm$ 0.004 & & 0.262 $\pm$ 0.004 & & 0.002 $\pm$ 0.001 & & 0.998 $\pm$ 0.001 & & 0.007 $\pm$ 0.000 & & 0.993 $\pm$ 0.000 & & 0.000 $\pm$ 0.000 & & 1.000 $\pm$ 0.000 \\
DDoS UDP & & 0.646 $\pm$ 0.010 & & 0.179 $\pm$ 0.007 & & 0.176 $\pm$ 0.008 & & 0.772 $\pm$ 0.036 & & 0.228 $\pm$ 0.036 & & 0.679 $\pm$ 0.012 & & 0.321 $\pm$ 0.012 & & 0.749 $\pm$ 0.042 & & 0.251 $\pm$ 0.042 \\
DoS ICMP & & 0.003 $\pm$ 0.001 & & 0.905 $\pm$ 0.005 & & 0.092 $\pm$ 0.005 & & 0.012 $\pm$ 0.004 & & 0.988 $\pm$ 0.004 & & 0.009 $\pm$ 0.001 & & 0.991 $\pm$ 0.001 & & 0.032 $\pm$ 0.009 & & 0.968 $\pm$ 0.009 \\
DoS SYN & & 0.205 $\pm$ 0.043 & & 0.247 $\pm$ 0.040 & & 0.547 $\pm$ 0.032 & & 0.442 $\pm$ 0.064 & & 0.558 $\pm$ 0.064 & & 0.791 $\pm$ 0.041 & & 0.209 $\pm$ 0.041 & & 0.885 $\pm$ 0.073 & & 0.115 $\pm$ 0.073 \\
DoS TCP & & 0.039 $\pm$ 0.001 & & 0.845 $\pm$ 0.004 & & 0.116 $\pm$ 0.003 & & 0.036 $\pm$ 0.010 & & 0.964 $\pm$ 0.010 & & 0.042 $\pm$ 0.001 & & 0.958 $\pm$ 0.001 & & 0.043 $\pm$ 0.002 & & 0.957 $\pm$ 0.002 \\
DoS UDP & & 0.338 $\pm$ 0.023 & & 0.556 $\pm$ 0.023 & & 0.106 $\pm$ 0.011 & & 0.142 $\pm$ 0.010 & & 0.858 $\pm$ 0.010 & & 0.388 $\pm$ 0.022 & & 0.612 $\pm$ 0.022 & & 0.367 $\pm$ 0.023 & & 0.633 $\pm$ 0.023 \\
\hline
\end{tabular}
}
\end{sideways}

%% file: tab7.tex
\renewcommand{\arraystretch}{1.6}
\centering
\caption{Classification of unknown attacks using Zhang \textit{et al} \cite{Zhang2021} CICIDS2017 dataset}
\resizebox{\columnwidth}{!}{%
\label{tab:unk2}
\begin{tabular}{l c c c c c c c c c c}
    \hline
     Unknown attack & & \multicolumn{4}{c}{AUROC} & &
     \multicolumn{4}{c}{AUPRC}\\
     \cline{3-6}
     \cline{8-11}
     & & EFC & Baseline &  ODIN \cite{Liang2018} &  OCN \cite{Zhang2021} & & EFC & Baseline & ODIN \cite{Liang2018} & OCN \cite{Zhang2021}\\
     \hline
     \textit{DoS slowloris} & & 0.693 & 0.172 & 0.718 & \textbf{0.926} & & \textbf{0.983} & 0.541 & 0.668 & 0.924\\
     \textit{Web Attack} & & 0.830 & 0.467 & 0.847 & \textbf{0.924} & &  \textbf{0.991} & 0.960 & 0.832 & 0.923\\
     \textit{Heartbleed}& & 0.990 & 0.96 & 0.881 & \textbf{1.000} &  & \textbf{1.000} & 0.762 & 0.885 & \textbf{1.000} \\
     \textit{DoS Slowhttp} & & 0.658 & 0.093 & 0.473 & \textbf{0.836} &  & \textbf{0.987} & 0.517 & 0.498 & 0.806\\
     \textit{Infiltration} & & \textbf{0.993} & 0.499 & 0.778 & 0.967 &  & \textbf{1.000} & 0.660 & 0.814 & 0.976 \\
     \textit{Botnet} & & \textbf{0.992} & 0.447 & 0.869 & 0.974 &  & \textbf{1.000} & 0.638 & 0.887 & 0.980\\
     \hline
     \textbf{Average} & & 0.859 & 0.397 & 0.761 & \textbf{0.938} &  & \textbf{0.993} & 0.628 & 0.764 & 0.935\\
     \hline
\end{tabular}
}

%% file: tab8.tex
\centering
\caption{Classification of unknown attacks using Dadkhah \textit{et al} \cite{CICIoMT2024} CICIoMT2024}
\resizebox{0.75\columnwidth}{!}{%
\label{tab:unk2-2024}
\centering
\begin{tabular}{l c c c c c c c c }
    \hline
     Unknown attack & & \multicolumn{3}{c}{AUROC} & &
     \multicolumn{3}{c}{AUPRC}\\
     \cline{3-5}
     \cline{7-9}
     & & EFC & & OCN \cite{Zhang2021} & & EFC & & OCN \cite{Zhang2021}\\
     \hline
\textit{ARP Spoofing} & & \textbf{0.602} & & 0.449 & & 0.231 & & \textbf{0.438} \\
\textit{DDoS Connect Flood} & & 0.877 & & \textbf{0.941} & & 0.544 & & \textbf{0.857} \\
\textit{DDoS ICMP} & & \textbf{0.790} & & 0.635 & & 0.432 & & \textbf{0.520} \\
\textit{DDoS Publish Flood} & & 0.876 & & \textbf{0.987} & & 0.541 & & \textbf{0.979} \\
\textit{DDoS SYN} & & \textbf{0.871} & & 0.720 & & 0.537 & & \textbf{0.623} \\
\textit{DDoS TCP} & & 0.878 & & \textbf{0.935} & & 0.540 & & \textbf{0.831} \\
\textit{DDoS UDP} & & \textbf{0.833} & & 0.460 & & \textbf{0.484} & & 0.417 \\
\textit{DoS Connect Flood} & & 0.877 & & \textbf{0.962} & & 0.540 & & \textbf{0.920} \\
\textit{DoS ICMP} & & \textbf{0.900} & & 0.637 & & 0.559 & & \textbf{0.562} \\
\textit{DoS Publish Flood} & & 0.836 & & \textbf{0.943} & & 0.489 & & \textbf{0.846} \\
\textit{DoS SYN} & & \textbf{0.873} & & 0.872 & & 0.539 & & \textbf{0.800} \\
\textit{DoS TCP} & & \textbf{0.874} & & 0.617 & & \textbf{0.540} & & 0.500 \\
\textit{DoS UDP} & & \textbf{0.853} & & 0.802 & & 0.505 & & \textbf{0.726} \\
\textit{Malformed Data} & & 0.483 & & \textbf{0.907} & & 0.476 & & \textbf{0.835} \\
\textit{OS Scan} & & 0.548 & & \textbf{0.631} & & 0.197 & & \textbf{0.628} \\
\textit{Ping Sweep} & & 0.776 & & \textbf{0.855} & & 0.415 & & \textbf{0.798} \\
\textit{Port Scan} & & 0.525 & & \textbf{0.675} & & 0.168 & & \textbf{0.707} \\
\textit{VulScan} & & \textbf{0.609} & & 0.543 & & 0.232 & & \textbf{0.534} \\
\hline
\textbf{Average} & & \textbf{0.771} & & 0.754 & & 0.443 & & \textbf{0.696} \\
\hline
\end{tabular}}

%% file: journal.bbl
\begin{thebibliography}{10}
\expandafter\ifx\csname url\endcsname\relax
  \def\url#1{\texttt{#1}}\fi
\expandafter\ifx\csname urlprefix\endcsname\relax\def\urlprefix{URL }\fi
\expandafter\ifx\csname href\endcsname\relax
  \def\href#1#2{#2} \def\path#1{#1}\fi

\bibitem{ENISA}
M.~Anisetti, C.~Ardagna, M.~Cremonini, E.~Damiani, J.~Sessa, L.~Costa, \href{https://www.concordia-h2020.eu/wp-content/uploads/2021/03/White_paper_SecurityThreats.pdf}{\capitalisewords{Security threat landscape}}, Tech. rep., "Universit\`a degli studi di Milano and Telecom Italia", accessed: 2025-02-13 (2020).
\newline\urlprefix\url{https://www.concordia-h2020.eu/wp-content/uploads/2021/03/White_paper_SecurityThreats.pdf}

\bibitem{DCMS}
M.~Department~for Digital, Culture, S.~from United~Kingdon, \href{https://doc.ukdataservice.ac.uk/doc/8970/mrdoc/pdf/8970_csbs_2022_technical_annex.pdf}{\capitalisewords{Cyber security breaches survey 2022}}, Tech. rep., "Ipsos", accessed: 2025-02-13 (2022).
\newline\urlprefix\url{https://doc.ukdataservice.ac.uk/doc/8970/mrdoc/pdf/8970_csbs_2022_technical_annex.pdf}

\bibitem{ACSC}
A.~S. Directorate, \href{https://www.cyber.gov.au/sites/default/files/2024-11/asd-cyber-threat-report-2024.pdf}{\capitalisewords{2023–24 Annual Cyber Threat Report}}, Tech. rep., "Australian Cyber Security Centre (ACSC)", accessed: 2025-02-13 (2024).
\newline\urlprefix\url{https://www.cyber.gov.au/sites/default/files/2024-11/asd-cyber-threat-report-2024.pdf}

\bibitem{Tidjon2019}
L.~Tidjon, M.~Frappier, A.~Mammar, \capitalisewords{Intrusion Detection Systems: A Cross-Domain Overview}, IEEE Communications Surveys and Tutorials 21~(4) (2019) 3639--3681.
\newblock \href {https://doi.org/10.1109/COMST.2019.2922584} {\path{doi:10.1109/COMST.2019.2922584}}.

\bibitem{AHMAD2021102122}
A.~Ahmad, S.~B. Maynard, K.~C. Desouza, J.~Kotsias, M.~T. Whitty, R.~L. Baskerville, \capitalisewords{How can organizations develop situation awareness for incident response: A case study of management practice}, Computers \& Security 21~(4) (2021) 3639--3681.
\newblock \href {https://doi.org/10.1016/j.cose.2020.102122} {\path{doi:10.1016/j.cose.2020.102122}}.

\bibitem{Zhang2021}
Z.~Zhang, Y.~Zhang, D.~Guo, M.~Song, \capitalisewords{a scalable Network Intrusion Detection System Towards detecting, discovering, and learning unknown attacks}, International Journal of Machine Learning and Cybernetics 12~(6) (2021) 1649--1665.
\newblock \href {https://doi.org/10.1007/s13042-020-01264-7} {\path{doi:10.1007/s13042-020-01264-7}}.

\bibitem{apruzzese2022cross}
G.~Apruzzese, L.~Pajola, M.~Conti, \capitalisewords{The Cross-evaluation of Machine Learning-based Network Intrusion Detection Systems}, IEEE Transactions on Network and Service Management 19~(4) (2022) 5152--5169.
\newblock \href {https://doi.org/10.1109/TNSM.2022.3157344} {\path{doi:10.1109/TNSM.2022.3157344}}.

\bibitem{Buczak2016}
A.~L. Buczak, E.~Guven, \capitalisewords{A Survey of Data Mining and Machine Learning Methods for Cyber Security Intrusion Detection}, IEEE Communications Surveys and Tutorials 18~(2) (2016) 1153--1176.
\newblock \href {https://doi.org/10.1109/COMST.2015.2494502} {\path{doi:10.1109/COMST.2015.2494502}}.

\bibitem{pontes2019new}
C.~F.~T. Pontes, M.~M.~C. de~Souza, J.~J.~C. Gondim, M.~Bishop, M.~A. Marotta, \capitalisewords{A New Method for Flow-Based Network Intrusion Detection Using the Inverse Potts Model}, IEEE Transactions on Network and Service Management 18~(2) (2021) 1125--1136.
\newblock \href {https://doi.org/10.1109/TNSM.2021.3075503} {\path{doi:10.1109/TNSM.2021.3075503}}.

\bibitem{Liang2018}
S.~Liang, Y.~Li, R.~Srikant, \capitalisewords{Enhancing the reliability of out-of-distribution image detection in neural networks}, in: International Conference on Learning Representations, 2018, pp. 1--15.
\newblock \href {https://doi.org/10.48550/arXiv.1706.02690} {\path{doi:10.48550/arXiv.1706.02690}}.

\bibitem{CICIoMT2024}
S.~Dadkhah, E.~C.~P. Neto, R.~Ferreira, R.~C. Molokwu, S.~Sadeghi, A.~A. Ghorbani, \capitalisewords{CICIoMT2024: A benchmark dataset for multi-protocol security assessment in IoMT}, Internet of Things 28 (2024) 101351.
\newblock \href {https://doi.org/10.1016/j.iot.2024.101351} {\path{doi:10.1016/j.iot.2024.101351}}.

\bibitem{Wu1982}
F.-Y. Wu, \capitalisewords{The Potts Model}, Reviews of modern physics 54~(1) (1982) 235.
\newblock \href {https://doi.org/10.1103/RevModPhys.54.235} {\path{doi:10.1103/RevModPhys.54.235}}.

\bibitem{openset}
W.~J. Scheirer, A.~de~Rezende~Rocha, A.~Sapkota, T.~E. Boult, \capitalisewords{Toward Open Set Recognition}, IEEE Transactions on Pattern Analysis and Machine Intelligence 35~(7) (2013) 1757--1772.
\newblock \href {https://doi.org/10.1109/TPAMI.2012.256} {\path{doi:10.1109/TPAMI.2012.256}}.

\bibitem{Hendrycks2017}
D.~Hendrycks, K.~Gimpel, \capitalisewords{A baseline for detecting misclassified and out-of-distribution examples in neural networks}, 5th International Conference on Learning Representations, ICLR 2017 - Conference Track Proceedings (2017).
\newblock \href {https://doi.org/10.48550/arXiv.1610.02136} {\path{doi:10.48550/arXiv.1610.02136}}.

\bibitem{Wang2023svm}
W.~Wang, X.~Du, D.~Shan, R.~Qin, N.~Wang, \capitalisewords{Cloud Intrusion Detection Method Based on Stacked Contractive Auto-Encoder and Support Vector Machine}, IEEE Transactions on Cloud Computing 10~(3) (2022) 1634--1646.
\newblock \href {https://doi.org/10.1109/TCC.2020.3001017} {\path{doi:10.1109/TCC.2020.3001017}}.

\bibitem{Anyanwu2023}
G.~Anyanwu, C.~Nwakanma, J.~M. Lee, D.-S. Kim, \capitalisewords{RBF-SVM kernel-based model for detecting DDoS attacks in SDN integrated vehicular network}, Ad Hoc Networks 140 (2023) 103026.
\newblock \href {https://doi.org/10.1016/j.adhoc.2022.103026} {\path{doi:10.1016/j.adhoc.2022.103026}}.

\bibitem{Louk2023}
M.~H.~L. Louk, B.~A. Tama, \capitalisewords{Dual-IDS: A bagging-based gradient boosting decision tree model for network anomaly intrusion detection system}, Expert Systems with Applications 213 (2023) 119030.
\newblock \href {https://doi.org/10.1016/j.eswa.2022.119030} {\path{doi:10.1016/j.eswa.2022.119030}}.

\bibitem{Mughaid2023}
A.~Mughaid, S.~AlZu’bi, A.~Alnajjar, E.~AbuElsoud, S.~E. Salhi, B.~Igried, L.~Abualigah, \capitalisewords{Improved dropping attacks detecting system in 5g networks using machine learning and deep learning approaches}, Multimedia Tools and Applications 82~(9) (2023) 13973--13995.
\newblock \href {https://doi.org/10.1007/s11042-022-13914-9} {\path{doi:10.1007/s11042-022-13914-9}}.

\bibitem{Wang2023}
K.~Wang, A.~Zhang, H.~Sun, B.~Wang, \capitalisewords{Analysis of Recent Deep-Learning-Based Intrusion Detection Methods for In-Vehicle Network}, IEEE Transactions on Intelligent Transportation Systems (2023).
\newblock \href {https://doi.org/10.1109/TITS.2022.3222486} {\path{doi:10.1109/TITS.2022.3222486}}.

\bibitem{Thakkar2023}
A.~Thakkar, R.~Lohiya, \capitalisewords{Fusion of statistical importance for feature selection in Deep Neural Network-based Intrusion Detection System}, Information Fusion 24~(2) (2023) 1843--1854.
\newblock \href {https://doi.org/10.1016/j.inffus.2022.09.026} {\path{doi:10.1016/j.inffus.2022.09.026}}.

\bibitem{Wu2023}
Y.~Wu, L.~Nie, S.~Wang, Z.~Ning, S.~Li, \capitalisewords{Intelligent Intrusion Detection for Internet of Things Security: A Deep Convolutional Generative Adversarial Network-Enabled Approach}, IEEE Internet of Things Journal 10~(4) (2023) 3094--3106.
\newblock \href {https://doi.org/10.1109/JIOT.2021.3112159} {\path{doi:10.1109/JIOT.2021.3112159}}.

\bibitem{Al-Yaseen2017}
W.~L. Al-Yaseen, Z.~A. Othman, M.~Z.~A. Nazri, \capitalisewords{Multi-level hybrid support vector machine and extreme learning machine based on modified K-means for intrusion detection system}, Expert Systems with Applications 67 (2017) 296--303.
\newblock \href {https://doi.org/10.1016/j.eswa.2016.09.041} {\path{doi:10.1016/j.eswa.2016.09.041}}.

\bibitem{Cruz2017}
S.~Cruz, C.~Coleman, E.~M. Rudd, T.~E. Boult, \capitalisewords{Open set intrusion recognition for fine-grained attack categorization}, 2017 IEEE International Symposium on Technologies for Homeland Security, HST 2017 (2017) 1--6.\href {https://doi.org/10.1109/THS.2017.7943467} {\path{doi:10.1109/THS.2017.7943467}}.

\bibitem{Henrydoss2017}
J.~Henrydoss, S.~Cruz, E.~M. Rudd, M.~Gonther, T.~E. Boult, \capitalisewords{Incremental open set intrusion recognition using extreme value machine}, Proceedings - 16th IEEE International Conference on Machine Learning and Applications (ICMLA 2017) (2017) 1089--1093.\href {https://doi.org/10.1109/ICMLA.2017.000-3} {\path{doi:10.1109/ICMLA.2017.000-3}}.

\bibitem{Rudd2018}
E.~M. Rudd, L.~P. Jain, W.~J. Scheirer, T.~E. Boult, \capitalisewords{The Extreme Value Machine}, IEEE Transactions on Pattern Analysis and Machine Intelligence 40~(3) (2018) 762--768.
\newblock \href {https://doi.org/10.1109/TPAMI.2017.2707495} {\path{doi:10.1109/TPAMI.2017.2707495}}.

\bibitem{Yao2019}
H.~Yao, Q.~Wang, L.~Wang, P.~Zhang, M.~Li, Y.~Liu, \capitalisewords{An Intrusion Detection Framework Based on Hybrid Multi-Level Data Mining}, International Journal of Parallel Programming 47 (2019) 740--758.
\newblock \href {https://doi.org/10.1007/s10766-017-0537-7} {\path{doi:10.1007/s10766-017-0537-7}}.

\bibitem{korba}
A.~A. Korba, A.~Boualouache, Y.~Ghamri-Doudane, \capitalisewords{Zero-X: A Blockchain-Enabled Open-Set Federated Learning Framework for Zero-Day Attack Detection in IoV}, IEEE Transactions on Vehicular Technology 73~(9) (2024) 12399--12414.
\newblock \href {https://doi.org/10.1109/TVT.2024.3385916} {\path{doi:10.1109/TVT.2024.3385916}}.

\bibitem{DeepQN}
S.~Yu, R.~Zhai, Y.~Shen, G.~Wu, H.~Zhang, S.~Yu, S.~Shen, \capitalisewords{Deep Q-Network-Based Open-Set Intrusion Detection Solution for Industrial Internet of Things}, IEEE Internet of Things Journal 11~(7) (2024) 12536--12550.
\newblock \href {https://doi.org/10.1109/JIOT.2023.3333903} {\path{doi:10.1109/JIOT.2023.3333903}}.

\bibitem{Dandelion}
J.~Wu, H.~Dai, K.~B. Kent, J.~Yen, C.~Xu, Y.~Wang, \capitalisewords{Open Set Dandelion Network for IoT Intrusion Detection}, ACM Transactions on Internet Technologies 24~(1) (2024) 1--26.
\newblock \href {https://doi.org/10.1145/3639822} {\path{doi:10.1145/3639822}}.

\bibitem{Jin2023}
D.~Jin, S.~Chen, H.~He, X.~Jiang, S.~Cheng, J.~Yang, \capitalisewords{Federated Incremental Learning based Evolvable Intrusion Detection System for Zero-Day Attacks}, IEEE Network 37~(1) (2023) 125--132.
\newblock \href {https://doi.org/10.1109/MNET.018.2200349} {\path{doi:10.1109/MNET.018.2200349}}.

\bibitem{Rahman2020}
S.~A. Rahman, H.~Tout, C.~Talhi, A.~Mourad, \capitalisewords{Internet of Things Intrusion Detection: Centralized, On-Device, or Federated Learning?}, IEEE Network 34~(6) (2020) 310--317.
\newblock \href {https://doi.org/10.1109/MNET.011.2000286} {\path{doi:10.1109/MNET.011.2000286}}.

\bibitem{Sharafaldin2018}
I.~Sharafaldin, A.~H. Lashkari, A.~A. Ghorbani, \capitalisewords{Toward generating a new intrusion detection dataset and intrusion traffic characterization}, Proceedings of the 4th International Conference on Information Systems Security and Privacy 1 ({2018}) 108--116.
\newblock \href {https://doi.org/10.5220/0006639801080116} {\path{doi:10.5220/0006639801080116}}.

\bibitem{Jaynes1957}
E.~T. Jaynes, \capitalisewords{Information theory and statistical mechanics. II}, Physical Review 108~(2) (1957) 171.
\newblock \href {https://doi.org/10.1103/PhysRev.108.171} {\path{doi:10.1103/PhysRev.108.171}}.

\end{thebibliography}
